\documentclass[10pt, conference, letterpaper]{IEEEtran}
\IEEEoverridecommandlockouts
\usepackage{amsmath,amssymb,amsfonts}
\usepackage{algorithmic}
\usepackage{algorithm}
\usepackage{graphicx}
\usepackage{textcomp}
\usepackage{xcolor}
\usepackage{amsmath}  % For mathematical equations
\usepackage{adjustbox}
\usepackage{relsize}
\usepackage{caption}
\usepackage{subfig}
\usepackage{url}
\usepackage{dsfont}
\usepackage{subcaption}
\def\BibTeX{{\rm B\kern-.05em{\sc i\kern-.025em b}\kern-.08em
    T\kern-.1667em\lower.7ex\hbox{E}\kern-.125emX}}

\usepackage{xcolor}

\begin{document}

% \title{Enhancing Wireless Multipath Transmission with Transformer-based Agents}
%\title{Edge-Served Congestion Control for Wireless Multipath Transmission with a Transformer Agent}
%\title{Decoupled Transformer-Based Congestion Control for Multipath Wireless Networks}

\title{Transformer-Based Multipath Congestion Control: A Decoupled Approach for Wireless Uplinks}

% \author{
% \IEEEauthorblockN{Liang Wang
% % Liang Wang\thanks{Corresponding author: wangliang1@bjtu.edu.cn}
% \IEEEauthorblockA{
% \textit{School of Electronic and Information Engineering} \\
% \textit{Beijing Jiaotong University} \\
% Beijing, China \\
% Email: wangliang1@bjtu.edu.cn}
% }
\author{Zongyuan Zhang\textsuperscript{*}, Tianyang Duan\textsuperscript{*}, Liang Wang, Zihan Fang, Zheng Lin, Yijun Lu, Jiening Wu, Xia Du, \\Miao Yang,  Zhe Chen, Heming Cui,~\IEEEmembership{Member,~IEEE} and Jun Luo,~\IEEEmembership{Fellow,~IEEE}  %and Bernard D. Researcher$^{2}$% <-this % stops a space 

\thanks{Z. Zhang, T. Duan, and H. Cui are with the Division of Computer Science, The University of Hong Kong, Hong Kong SAR, China
(e-mail: tyduan@cs.hku.hk; zyzhang2@cs.hku.hk; heming@cs.hku.hk).}
\thanks{L. Wang is with the School of Electronic and Information Engineering, Beijing Jiaotong University, Beijing, China (e-mail: wangliang1@bjtu.edu.cn).}
\thanks{Z. Fang is with the Department of Computer Science, City University of Hong Kong, Kowloon, Hong Kong SAR, China (e-mail: zihanfang3-c@my.cityu.edu.hk).}
\thanks{Y. Lu is with the Department of Computer Science and Engineering, Waseda University, Tokyo,  Japan (e-mail: yijun@ruri.waseda.jp).}
\thanks{Z. Lin is with the Department of Electrical and Electronic Engineering, University of Hong Kong, Pok Fu Lam, Hong Kong SAR, China (e-mail: linzheng@eee.hku.hk).}
\thanks{J. Wu is with the College of Artificial Intelligence, Southwest University, Chongqing 400715, China (e-mail:  jnwu66@swu.edu.cn).}
\thanks{X. Du and M. Yang are with the School of Computer and Information
Engineering, Xiamen University of Technology, Xiamen, 361000, China
(email: duxia@xmut.edu.cn; yangmiao@xmut.edu.cn ).}
\thanks{Z. Chen is with the Institute of Space Internet, Fudan University, Shanghai 200438, China (e-mail: zhechen@fudan.edu.cn).}
\thanks{J. Luo is with the School of Computer Engineering, Nanyang Technological University, Singapore (e-mail: junluo@ntu.edu.sg).}
\thanks{\textit{(Corresponding author: Liang Wang and Zihan Fang)}}
\thanks{* These authors contributed equally.}
}
%\thanks{Digital Object Identifier (DOI): 10.1109/LRA.2024.3396092}
% }
% \IEEEauthorblockA{
% \textit{Anonymous Department} \\
% \textit{Anonymous Institute} \\
% Unknown, Region \\
% Email: AnonymousAuthor@XXX}
% }

\maketitle

\begin{abstract}
The proliferation of artificial intelligence applications on edge devices necessitates efficient transport protocols that leverage multi-homed connectivity across heterogeneous networks. While Multipath TCP enables bandwidth aggregation, its in-kernel congestion control mechanisms lack the programmability and flexibility needed for achieving efficient transmission. Additionally, inherent measurement noise renders network state partially observable, challenging data-driven approaches like deep reinforcement learning (DRL). To address these challenges, we propose a Transformer-based Congestion Control Optimization (TCCO) framework for multipath transport. TCCO employs a decoupled architecture that offloads control decisions to an external decision engine via a lightweight in-kernel client and user-space proxy, enabling edge devices to leverage external computational resources while maintaining TCP/IP compatibility. The Transformer-based DRL agent in the external decision engine uses self-attention to capture temporal dependencies, filter noise, and coordinate control across subflows through a unified policy. Extensive evaluation on both simulated and real dual-band Wi-Fi testbeds demonstrates that TCCO achieves superior adaptability and performance than state-of-the-art baselines, validating the feasibility and effectiveness of TCCO for wireless networks.
\end{abstract}


\section{Introduction}
\label{section:1}
As artificial intelligence (AI) applications~\cite{duan2025leed,zhang2025robust,hong2026conflict,fang2025dynamic,sun2025intra,lin2024adaptsfl,yuan2023graph} proliferate on edge devices, multi-homed devices equipped with heterogeneous network interfaces, such as Wi-Fi, LTE, and 5G, have become essential components of edge networks. These devices offload compute-intensive AI workloads to edge servers when local processing capacity is insufficient. For instance, in industrial manufacturing, multi-homed sensors and controllers transmit high-dimensional data streams to nearby edge nodes for real-time inference, exploiting multiple network paths to maintain uninterrupted communication despite individual link congestion or failure \cite{pokhrel2021multipath,silva2021iot}. By leveraging heterogeneous paths simultaneously, multi-homed devices can enhance transmission reliability and reduce latency, addressing the stringent quality-of-service requirements of latency-sensitive AI applications. 

Multipath transport protocols play a critical role in harnessing the full potential of multi-homed devices. Multipath TCP (MPTCP), standardized by the IETF in RFC 8684 \cite{rfc8684}, has emerged as the dominant solution, with its widespread adoption in Linux kernel and Apple iOS. The efficacy of MPTCP fundamentally depends on its congestion control (CC) mechanism, which balances throughput optimization with fairness by adjusting each subflow's sending rate in response to network conditions. To achieve optimal bandwidth utilization and controlled queueing depth, the CC mechanism must dynamically adapt to the diverse and time-varying characteristics of each subflow.

However, implementing adaptive CC algorithms within the kernel presents substantial obstacles \cite{narayan2018restructuring}. First, kernel development is challenging due to its stringent programming environment, characterized by limited debugging tools and complex concurrency management, which requires specialized expertise and lengthy development cycles. Second, the kernel presents a resource-constrained execution environment, ill-suited for sophisticated computational models that require features unavailable in kernel space, such as floating-point arithmetic, access to user-space libraries, and hardware accelerators like GPUs. Third, kernel-level implementations carry inherent risks, minor errors can trigger system-wide failures and critical security vulnerabilities.

These limitations have driven an industry-wide trend toward relocating network functions from the monolithic kernel to enhance datapath programmability. Kernel-bypass technologies such as DPDK \cite{intel_dpdk_2014} and user-space protocols like QUIC \cite{Iyengar2021QUIC} exemplify this shift. These approaches embody a broader architectural principle: decoupling complex, evolving control logic from the static, performance-critical packet forwarding plane. This separation creates opportunities for integrating adaptive, data-driven CC methodologies.

Among data-driven approaches, deep reinforcement learning (DRL) is promising for adaptive CC as it learn policies that map network observations to transmission decisions without explicit modeling of network dynamics \cite{Xiao2021,Giacomoni2024}. However, conventional single-step DRL methods rely on instantaneous state observations, which are inherently noisy due to ACK aggregation and transient channel fluctuations. This observation noise causes agents to react to transient artifacts rather than underlying network dynamics, leading to suboptimal control decisions. While extending the observation window can mitigate observation noise, it reduces responsiveness to genuine network variations. Single-step DRL formulations cannot satisfy these conflicting requirements. This fundamental challenge motivates our proposed approach, as further elaborated in Section~\ref{sec:motive}.

%To address these challenges, we propose Transformer-based Congestion Control Optimization (TCCO), a framework that enables adaptive CC through edge-based policy learning. First, to overcome the limitations of in-kernel CC implementation while maintaining compatibility with the existing TCP/IP stack, we decouple MPTCP CC logic from its in-kernel execution by implementing a lightweight kernel client that enforces directives from a decision engine operating outside the kernel. This architecture allows the system to leverage edge computing resources for real-time, data-driven CC. Second, to handle the inherent noise in network measurements and the complex inter-subflow dependencies, we employ a Transformer-based DRL agent as the decision engine. Unlike conventional DRL that process observations sequentially, the Transformer leverages self-attention mechanisms to jointly model temporal dependencies across historical observations, effectively filtering transient fluctuations and extracting robust network state representations. This enables the agent to infer underlying network dynamics and generate a coordinated control policy across all subflows. The contributions of this paper are as follows:
To address these challenges, we propose Transformer-based Congestion Control Optimization (TCCO), a framework that enables adaptive CC through edge-based policy learning. First, to overcome the limitations of in-kernel CC implementation while maintaining compatibility with the existing TCP/IP stack, we decouple MPTCP CC logic from its in-kernel execution. We implement a lightweight kernel client that communicates with an external decision engine via a user-space proxy, offloading sophisticated CC to external resources. This architecture allows TCCO to leverage edge computing resources for real-time, data-driven CC. Second, to handle the inherent noise in network measurements and the complex inter-subflow dependencies, we employ a Transformer-based DRL agent as the external decision engine. The Transformer-based DRL agent leverages self-attention mechanisms to jointly model temporal dependencies across historical observations, effectively filtering transient fluctuations and extracting robust network state representations. This enables TCCO to infer underlying network dynamics and generate a coordinated control policy across all subflows. The contributions of this paper are as follows:
\begin{itemize}
\item We propose TCCO, a Transformer-based congestion control framework deployed and validated in real-world network infrastructure, demonstrating seamless compatibility with existing protocol stacks and practical adaptive learning capabilities.
\item We design a decoupled architecture separating decision-making from the kernel datapath, enabling agile deployment of DRL-based CC strategies on edge infrastructure.
\item We develop a Transformer-based DRL agent that distinguishes network dynamics from transient noise, achieving robust multi-subflow control.
\item We empirically evaluate TCCO on simulated topologies and a real dual-band Wi-Fi testbed, and demonstrate that TCCO consistently outperforms state-of-the-art baselines across both local and edge deployments.
\end{itemize}

The rest of this paper is organized as follows. Section \ref{sec:motive} presents the challenge and motivation of the paper. Section \ref{sec:framework} details the TCCO framework design. Section \ref{sec:validation} provides performance evaluation results. Section \ref{sec:related} discusses related work, and Section \ref{sec:conclusion} concludes the paper.

\begin{figure}[t]
  \centering
  \includegraphics[width=\columnwidth]{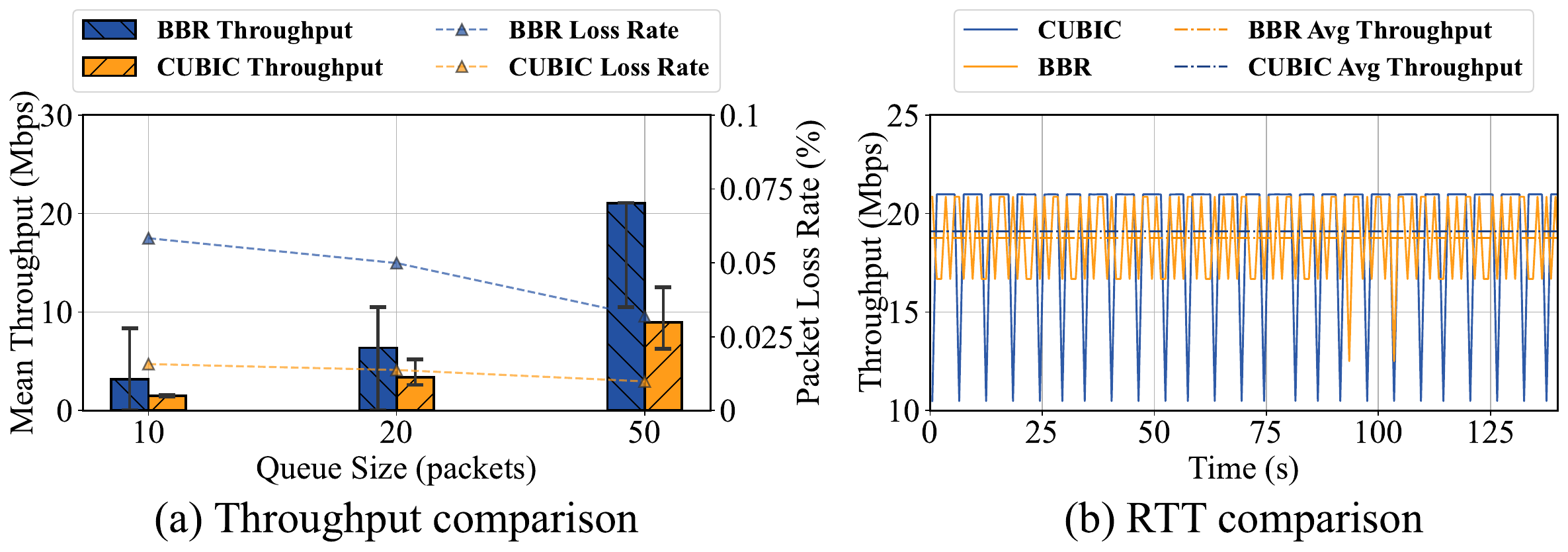}
  \caption
  {Performance comparison between CUBIC and BBR. The queue buffer was simulated using the Linux Traffic Control (TC). During the test in (b), the buffer size was set to 1000 packets (greater than the actual requirement).}
  \label{fig1}
  \end{figure}

\section{Challenge and Motivation}
\label{sec:motive}

In this section, we motivate our approach through empirical analysis of three key questions: first, the rationale for employing DRL-based methods; second, the partial observability problem that limits DRL; and third, how the Transformer overcomes these challenges for more robust CC. 

\subsection{From Traditional Algorithms to DRL-based Methods}
Traditional CC algorithms exhibit inherent limitations due to their reliance on hand-crafted heuristics with static parameter tuning. Such designs struggle to adapt to heterogeneous and time-varying wireless links, leading to inconsistent performance across diverse network conditions. To demonstrate this, we conducted experiments transferring 1GB of data between two hosts under two representative scenarios. In the first scenario (Fig.\ref{fig1}(a)), configured with a 200Mbps link, 60ms delay, 0.01\% packet loss, and a shallow buffer. BBR \cite{cardwell2016bbr} delivered higher throughput but at the cost of greater packet loss. In the second scenario (Fig.\ref{fig1}(b)), emulating a small Bandwidth-Delay Product (BDP) network (20Mbps, 5ms delay, sufficient buffer), CUBIC\cite{ha2008cubic} attained superior throughput yet showed significant fluctuations in both sending rate and RTT. These observations demonstrate that the heuristic-based algorithm's performance in terms of delivery rate, delay, and stability is highly dependent on the specific network environment.

To address these limitations, learning-based methods, particularly deep reinforcement learning (DRL), have emerged as a promising alternative. DRL enables agents to derive adaptive control policies through online interaction with networks, eliminating the need for explicit analytical models. Early work validated the feasibility of DRL-based transmission strategies: QTCP \cite{Li2018QTCP} applied Q-learning to congestion control, while PCC \cite{Dong2015PCC} employed online optimization to iteratively refine sending rates based on observed performance. DRL-based methods have also been extended to multipath transport, including delay-aware path scheduling \cite{l2mptcp}, multi-agent frameworks for inter-subflow coordination \cite{He2021DeepCC}, and multipath optimization for distributed edge learning \cite{fair_efficient_mptcp}.

%DRL provides a principled approach to addressing these challenges. By learning a unified control policy for the entire multipath connection through direct interaction with the environment, DRL inherently captures cross-path dependencies that are difficult to model analytically. Furthermore, unlike heuristic-based methods, DRL derives adaptive strategies from experience without requiring explicit network models, enabling generalization across heterogeneous conditions. The optimization objective is explicitly defined in a customizable reward function, allowing the learned policy to be tailored to diverse requirements, such as minimizing latency or maximizing aggregate throughput.
  
\subsection{Limitations of DRL-based Methods under Partial Observability}

\begin{figure}[t]
\centering
\includegraphics[width=\columnwidth]{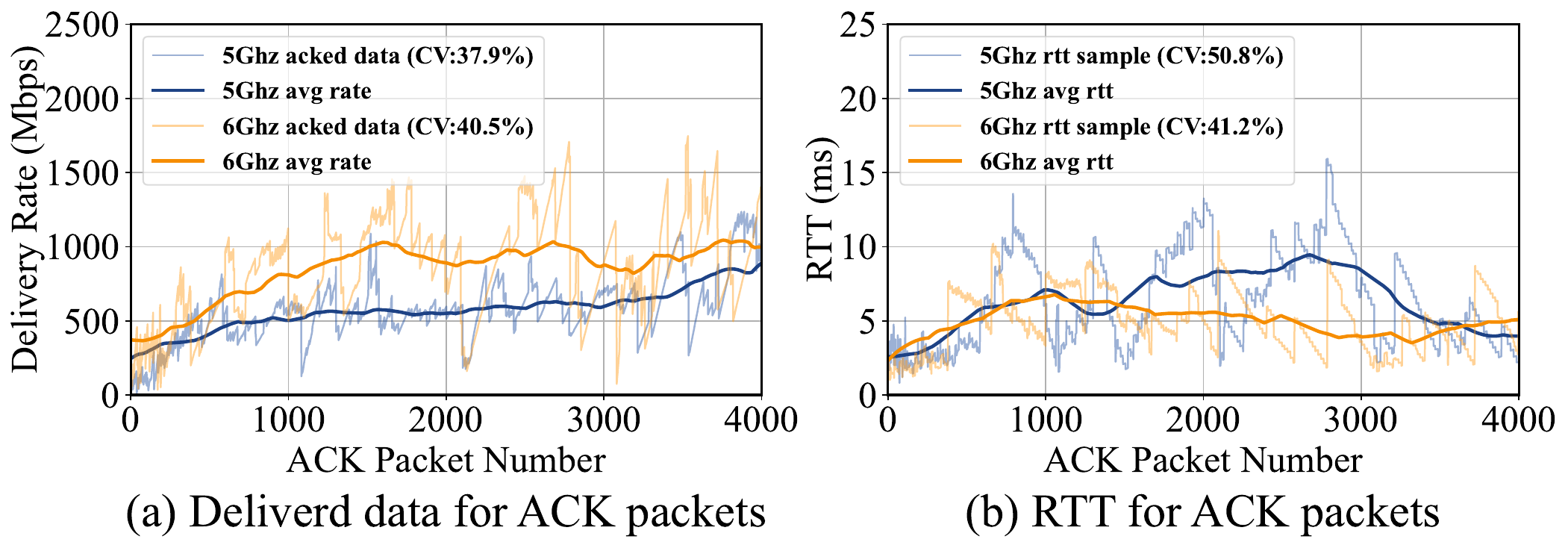}
\caption
{Statistics of ACK packets during a real MPTCP upload. The test was conducted on a dual-band, dual-link Wi-Fi network supporting 5GHz and 6GHz.}
\label{fig2}
\end{figure}

Although these DRL-based methods have achieved progress, they model CC as a fully observable Markov decision process, sharing a common limitation: relying on instantaneous input inherently results in inaccurate estimates of throughput and latency. A preliminary experiment on a dual-band Wi-Fi multipath network demonstrated this phenomenon: during an MPTCP upload using BBR (as shown in Fig. \ref{fig2}), both the 5 GHz and 6 GHz links exhibited high coefficients of variation (CV) in delivered data volume and RTT, with instantaneous rate estimates exceeding the true average by over 40\%. Such measurement variability fundamentally undermines the agent's ability to assess network conditions accurately and track state transitions reliably. This mismatch between high-frequency measurement noise and slower, true network dynamics means the agent operates with only partial visibility into the true network state. The control loop must react to sub-millisecond-level measurement noise, while the true network dynamics (e.g., changes in available bandwidth) evolve over longer timescales. Agents that cannot separate transient fluctuations from persistent trends based on a sequence of misleading observations will make erratic decisions, degrading CC performance.

A straightforward mitigation is to observe the network feedback over an extended window until conditions stabilize. However, longer aggregation periods delay responses to genuine state changes. In environments where network conditions shift on timescales comparable to or shorter than the observation window, the agent may miss critical transitions entirely and respond only after the network has evolved to a subsequent state. Furthermore, optimal window length varies across environments: a duration suitable for stable wired links can be inadequate for volatile wireless channels. This lack of generalizability undermines DRL's promise of adaptive policies without extensive manual tuning. 

Instead of fixed-window aggregation, we let the agent to process sequences of historical observations directly. This preserves fine-grained information while maintaining high decision frequency, but requires an architecture that can capture long-range temporal dependencies. The Transformer is well-suited for this task. The self-attention mechanism dynamically weighs different points in the observation sequence, allowing the agent to distinguish genuine network trends from transient noise. This enables proactive decisions based on underlying dynamics rather than reactive responses to noisy measurements.

\begin{figure}[t]
  \centering
  \includegraphics[width=\columnwidth]{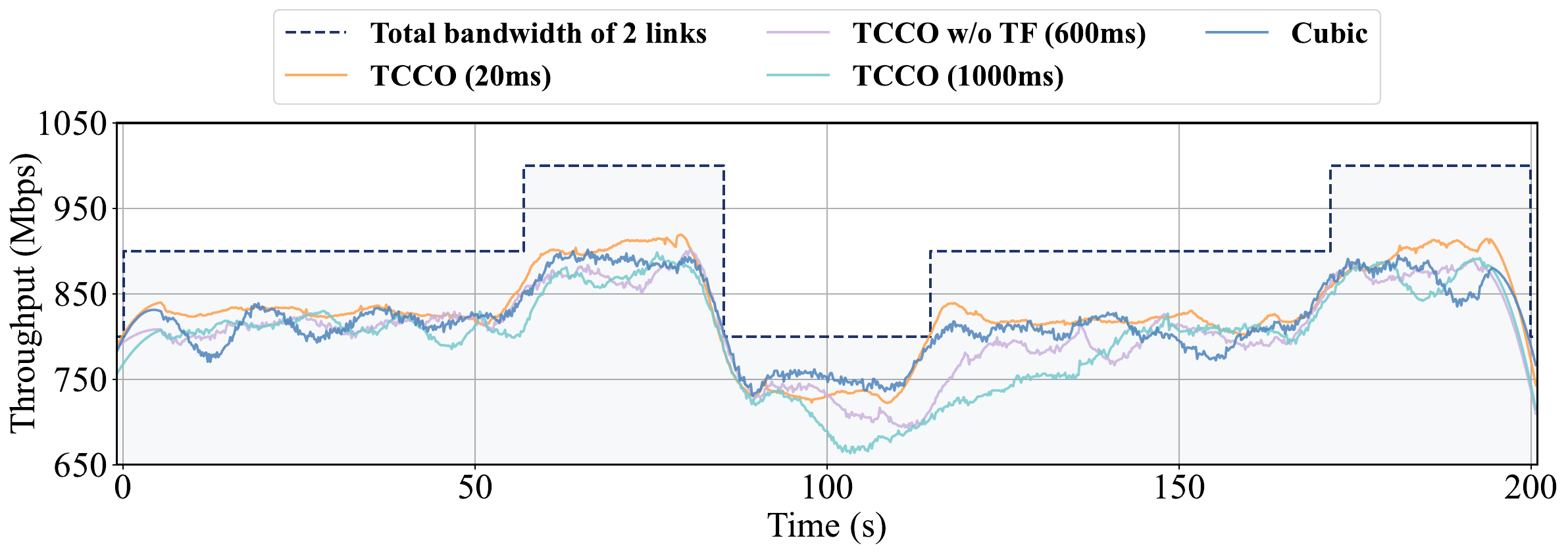}
  \caption
  {Responsiveness with different control intervals. TCCO (1000ms) uses artificially introduced delay. The dashed line shows the total bandwidth (altered via Linux TC) of paths.}
  \label{fig3}
  \end{figure}

We conducted experiments to evaluate the impact of decision interval on control performance. The evaluation was performed in an emulated multipath network with periodically fluctuating link capacity (bandwidth alternating between 400 and 500 Mbps, delay varying between 3 and 5 ms). As shown in Fig. \ref{fig3}, TCCO operating at a decision interval of 20 ms (including per-step training overhead), responded effectively to bandwidth variations and achieved an average throughput of 817 Mbps, outperforming CUBIC (814 Mbps) which showed slower adaptation to capacity changes. In contrast, TCCO without the Transformer component (TCCO w/o TF) required decision intervals on the order of hundreds of milliseconds to ensure stable model updates, which diminished its responsiveness to network dynamics (804 Mbps). Furthermore, extending the decision interval of TCCO to approximately 1000 ms resulted in significant performance degradation (775 Mbps), confirming that fine-grained temporal control is essential for maximizing throughput in dynamic network environments.

\section{TCCO Framework}
\label{sec:framework}
\subsection{System Overview}
In this section, we propose Transformer-based Congestion Control Optimization (TCCO), which adopts a decoupled architecture that separates decision-making logic from the in-kernel datapath, as illustrated in Fig.~\ref{fig4}. This design enables multi-homed edge devices, such as industrial sensors transmitting high-dimensional data streams over heterogeneous interfaces to leverage external computational resources for sophisticated congestion control while maintaining efficient packet processing. TCCO comprises three key components that form a closed-loop control system:

\begin{figure}[t]
  \centering
  \includegraphics[width=\columnwidth]{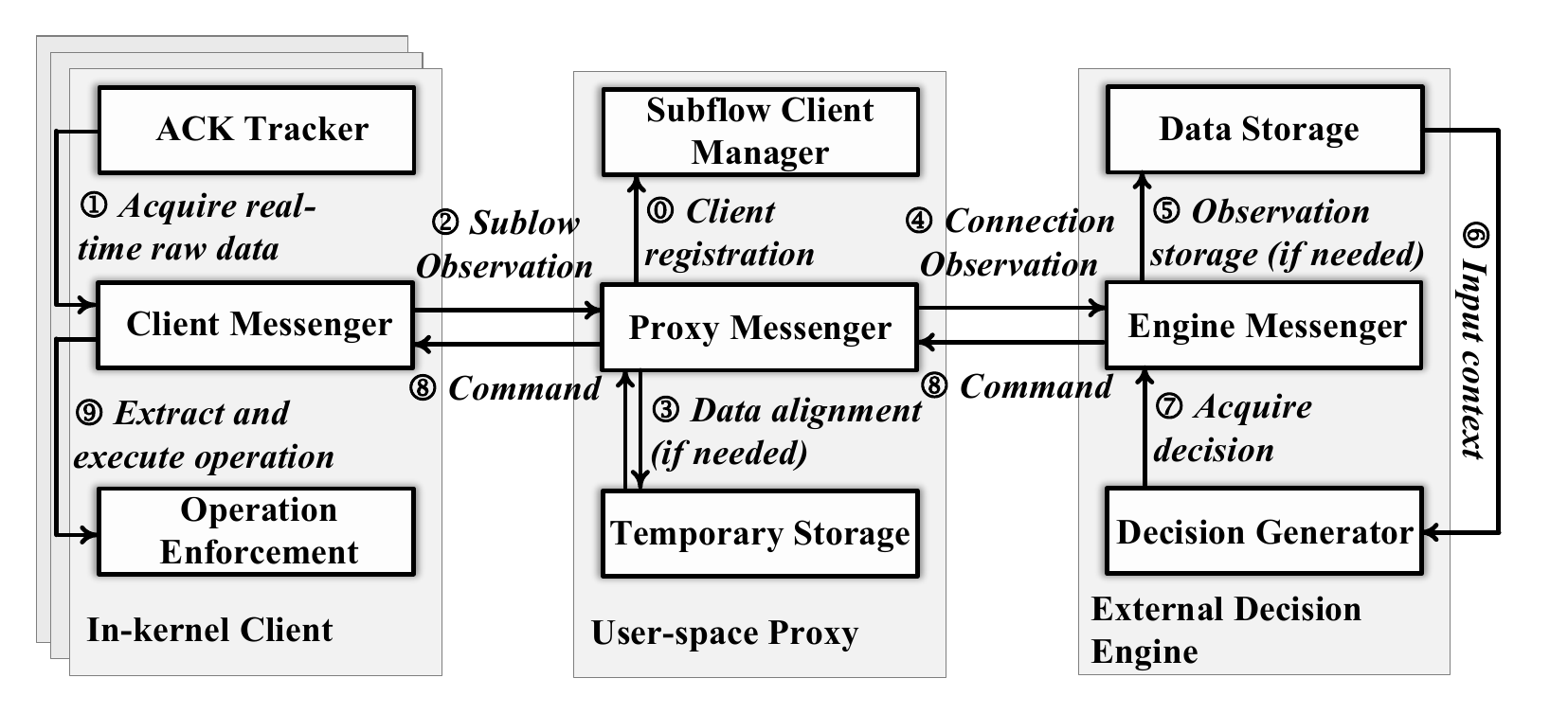} 
  \caption{The concept of network function decoupling in TCCO. The external decision engine can operate in edge devices. Communication methods between modules are flexible. }
  \label{fig4}
\end{figure}

\textbf{In-kernel Client.}
The in-kernel client interacts with the datapath via lightweight hooks to perform three coordinated tasks. First, it monitors transmission status by retrieving precise metrics from kernel data structures, including both instantaneous values such as the latest RTT and aggregated statistics such as the maximum RTT. Second, upon trigger events like ACK arrivals, it serializes metrics and exports them to user space. Third, it receives control directives from user space and translates them into atomic kernel operations that adjust transmission parameters. All interactions remain thread-safe and lock-free to preserve normal kernel functionality.

\textbf{User-space Proxy.}
The proxy serves as an intermediary between in-kernel clients and the external decision engine, managing dedicated clients for each subflow in multipath scenarios. Its invocation controller determines when to trigger decision generation, optionally waiting for metric reports from all active clients before initiating coordinated multi-flow optimization.

\textbf{External Decision Engine.}
The decision engine transforms aggregated metrics into structured state representations and computes CC strategies by a Transformer-based DRL agent. To enhance flexibility and scalability, it can be offloaded to edge devices or dedicated servers. The engine also maintains persistent storage for historical data, enabling decisions informed by long-term traffic patterns.

\subsection{System Model}

\begin{figure*}[t]
  \centering
  \includegraphics[width=0.98\textwidth]{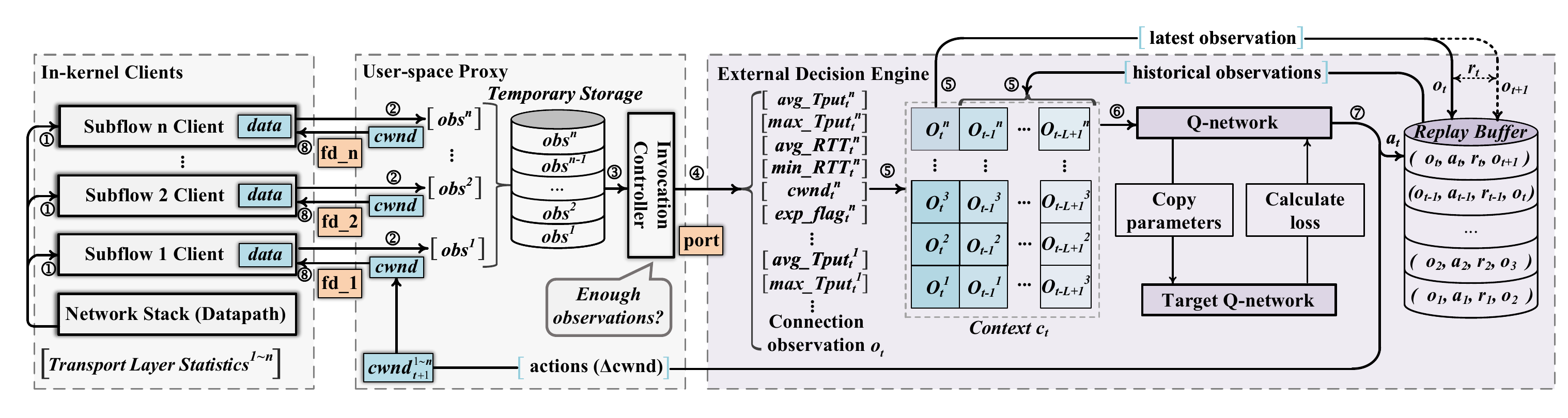}
  \caption{An overview of TCCO framework.}
  \label{fig5}
\end{figure*}

\label{sec:problem}
\subsubsection{Cross-Subflow Coupling Dynamics}
We begin by analyzing the scheduling-induced coupling mechanism that arises when multiple subflows share a common scheduler. We conduct our analysis based on the default \textit{minRTT} scheduler~\cite{wu2021multipath}; analogous coupling patterns can be observed under alternative schedulers as well.

Consider a multipath connection maintaining $M$ parallel subflows distributed across a set of network paths $\mathcal{W} = \{1, 2, \ldots, M\}$. The multipath scheduler introduces implicit coupling among subflows through its packet-to-subflow assignment logic. To formalize this coupling mechanism, we first define the subflow availability indicator:
\begin{equation}
\label{phi}
\phi_i(t) = \mathbb{I}[cwnd_i(t) > (q_i(t) + f_i(t)) \wedge mode_i(t) \neq recovery]
\end{equation}
where $cwnd_i(t)$ represents the congestion window size of subflow $i$, which limits the maximum amount of unacknowledged data permitted on that subflow. $q_i(t)$ denotes the number of packets currently queued in the send buffer awaiting transmission on subflow $i$, and $f_i(t)$ denotes the number of packets that have been transmitted but not yet acknowledged (i.e., in-flight packets). The condition $mode_i(t) \neq recovery$ ensures that the subflow is not currently executing loss recovery procedures, during which it should not accept new data. A subflow is deemed available for scheduling only when its congestion window has sufficient capacity to accommodate additional packets and it operates in normal transmission mode.

Given the set of available subflows, the scheduler assigns each incoming packet to the subflow exhibiting the minimum round-trip time:
\begin{equation}
\label{jstar}
j^*(t) = \arg\min_{i: \phi_i(t)=1} RTT_i(t)
\end{equation}
where $j^*(t)$ identifies the index of the selected subflow at time $t$, and $RTT_i(t)$ denotes the current round-trip time estimate for subflow $i$. This selection policy reflects the intuition that routing traffic through lower-latency paths improves overall connection performance, though as we demonstrate subsequently, it simultaneously introduces coupling asymmetry and cross-flow blindness.

\textbf{Property 1 (Coupling Asymmetry).} When subflow $i$ operates with an excessive congestion window $cwnd_i$, queuing delay accumulates, causing the total round-trip time to decompose as:
\begin{equation}
RTT_i(t) = RTT_{\min,i}(t) + Q_{queue,i}(t)
\end{equation}
where $RTT_{\min,i}(t)$ represents the minimum achievable RTT (i.e., the propagation delay) and $Q_{queue,i}(t)$ denotes the queuing delay experienced by subflow $i$ at time step $t$. This elevated RTT causes the scheduler to favor alternative paths with lower latency. Conversely, when $cwnd_i$ is insufficient to support additional transmissions, the indicator becomes $\phi_i(t) = 0$, effectively excluding subflow $i$ from scheduling consideration. This behavior follows directly from the scheduler's selection rule and the congestion window constraint specified in Eqn. \ref{phi} and \ref{jstar}. As a consequence, the traffic allocation probability for subflow $i$, which captures the coupling effects among subflows, can be expressed as:
\begin{equation}
\label{prob_allocation}
P_{alloc}(j^*(t)=i) = \frac{\mathbb{I}[RTT_i(t) = \min_{j: \phi_j(t)=1} RTT_j(t)]}{\sum_{k=1}^M \phi_k(t)}
\end{equation}

This creates asymmetric dependencies, whereby the performance of subflow $j$ is influenced by the congestion window decisions of all other subflows. Specifically, the load assigned to each subflow is given by:
\begin{equation}
\label{lambda}
\lambda_i(t) = \Lambda(t) \cdot P_{alloc}(j^*(t)=i)
\end{equation}
where $\Lambda(t)$ denotes the total arrival rate at time step $t$, and $\lambda_i(t)$ denotes the traffic load allocated to subflow $i$. This coupling mechanism implies that each subflow's effective throughput depends not only on its own state but also on the performance and availability of other subflows.

\textbf{Property 2 (Cross-flow Blindness).} The coupling described above introduces cross-flow blindness at the subflow level. Let $x^i_t$ denote the true local state of subflow $i$ at time step $t$, which evolves according to:
\begin{equation}
x^i_{t+1} = P_i(x^i_{t}, \Delta cwnd_i(t) , \lambda_i(t))
\end{equation}
where $x^i_{t+1}$ denotes the true local state of subflow $i$ at the subsequent time step, $P_i(\cdot)$ denotes the local system dynamics of subflow $i$, and $\Delta cwnd_i(t)$ represents the local congestion window adjustment action taken by subflow $i$.

The cross-flow blindness arises because $\lambda_i(t)$ depends on $cwnd_j(t)$ for all $j \neq i$, creating hidden dependencies that are unobservable to local controllers. Specifically, the traffic assignment $\lambda_i(t)$ in Eqn. \ref{lambda} is determined by the global state comprising the congestion window vector $cwnd(t) = [cwnd_1(t), \ldots, cwnd_n(t)]$ and the round-trip time vector $RTT(t) = [RTT_1(t), \ldots, RTT_n(t)]$ across all subflows. However, path $i$ can only observe its local metrics $(cwnd_i(t), RTT_i(t))$ and cannot directly access $(cwnd_j(t), RTT_j(t))$ for $j \neq i$. Consequently, from an individual subflow's perspective, the true system dynamics remain partially hidden, as the influence of other subflows manifests only indirectly through changes in the local state.

\begin{figure*}[t]
  \centering
  \includegraphics[width=0.98\textwidth]{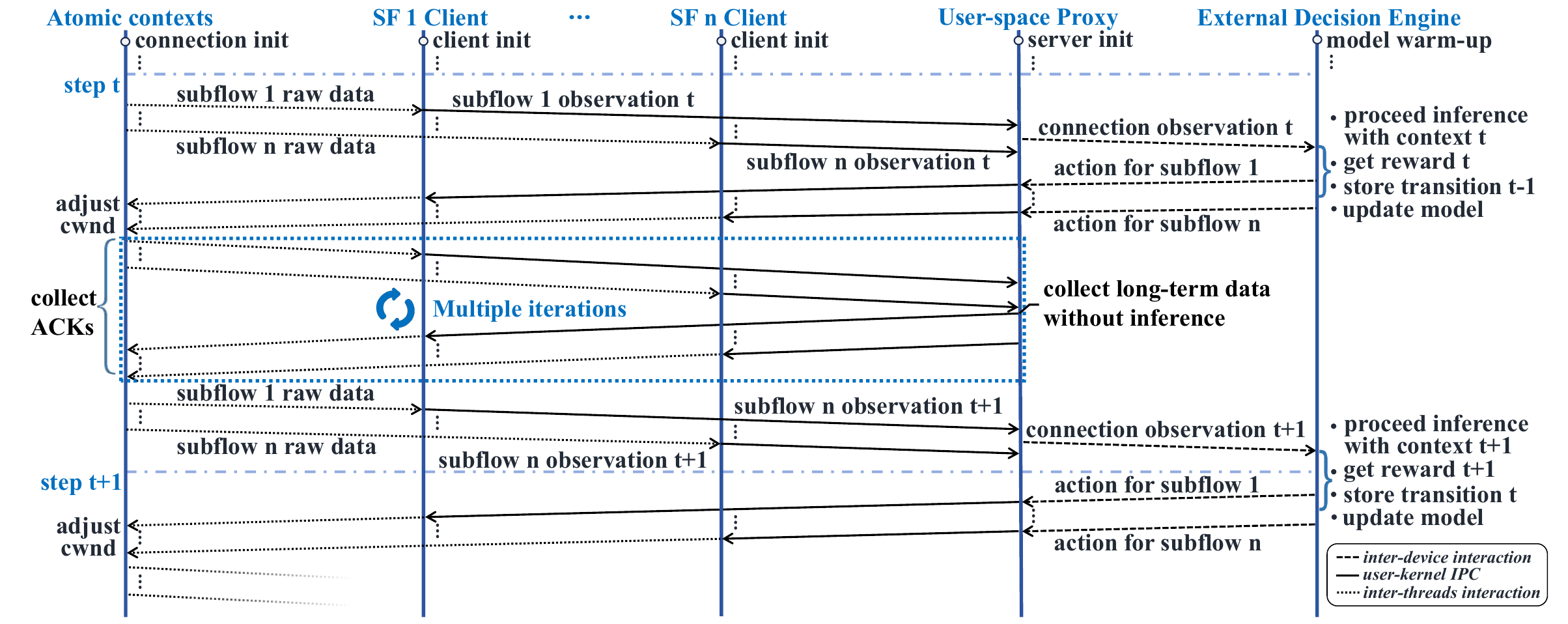}
  \caption{TCCO workflow and interactions.}
  \label{fig6}
  \end{figure*}
  
\subsubsection{POMDP Formulation for Multipath CC}
The cross-subflow coupling and cross-flow blindness established above reveal a fundamental limitation of independent single-flow control: each subflow optimizes based solely on its local observations while remaining unaware of the global state that governs traffic allocation. Consequently, the Nash equilibrium achieved by independent controllers is generally suboptimal compared to centralized joint control, as individual paths react to incomplete state information without accounting for the coupled dynamics induced by the scheduler. This observation motivates a joint optimization framework that explicitly models the partial observability inherent in multipath CC.

To this end, we formulate the multipath congestion control problem as a Partially Observable Markov Decision Process (POMDP) \cite{spaan2012partially}, defined by the tuple $\left \langle \mathcal{S}, \mathcal{A}, \mathcal{O}, P, R, \gamma \right \rangle $. The state space $\mathcal{S}$ comprises the complete information of all subflows, where each global state $s_t = \{x^i_t \mid i \in \mathcal{W}\}$ aggregates the local states across the subflow set $\mathcal{W}$. The action space $\mathcal{A}$ consists of joint congestion window adjustments $a_t = \{a^i_t \mid i \in \mathcal{W}\}$, where each $a^i_t = \Delta cwnd_i(t)$ specifies the window modification for subflow $i$. The observation space $\mathcal{O}$ reflects the cross-subflow blindness: each observation $o_t = \{o^i_t\}_{i \in \mathcal{W}}$ comprises locally observable metrics for each subflow and provides incomplete information about the true global state $s_t$. The state transition function $P: \mathcal{S} \times \mathcal{A} \rightarrow \mathcal{S}$ captures the coupled evolution of subflows through the scheduler traffic allocation:
\begin{equation}
\label{prob}
P(s_{t+1}|s_t, a_t) = \prod_{i=1}^M P_i(x^i_{t+1}|x^i_t, a^i_t, \lambda_i(t))
\end{equation}
where the dependence of $\lambda_i(t)$ on the global state, as established in Eqn.~\ref{lambda}, introduces the coupling among subflow transitions. The reward function $R: \mathcal{S} \times \mathcal{A} \rightarrow \mathbb{R}$ is designed to maximize total throughput while maintaining bounded latency across all subflows:
\begin{align}
\max \quad & \sum_{i=1}^M Tput_i(t) \\
\text{s.t.} \quad & \sum_{i=1}^M \max(0, RTT_i(t) - \tau_i) \leq \epsilon
\end{align}
where $Tput_i(t)$ denotes the time-averaged throughput of subflow $i$, $\tau_i$ denotes the minimum achievable RTT under zero queuing delay, and $\epsilon$ is the tolerable latency violation threshold. The discount factor $\gamma \in [0, 1)$ balances immediate and future rewards. %The objective is to find a policy $\pi: \mathcal{O} \rightarrow \mathcal{A}$ that maximizes the expected discounted cumulative reward $J(\pi) = \mathbb{E}_{\pi}\bigl[\sum_{t=0}^{\infty} \gamma^t r_t\bigr]$.

\subsection{System Design}
\label{sec:system}
TCCO decouples control logic from data-path operations to enable adaptive CC in multipath transmission. As illustrated in Fig.~\ref{fig5}, in-kernel clients handle per-subflow metric collection and congestion window enforcement, a user-space proxy aggregates time-aligned observations, and an external decision engine based on a Transformer-based DRL agent processes historical observation sequences to derive context-aware, cross-path control decisions.

\subsubsection{Observation Pipeline}

Each subflow is managed by a dedicated in-kernel client responsible for metric exposure and directive enforcement. These clients communicate with the decision engine through a user-space proxy that performs metric aggregation and command demultiplexing, as depicted in Fig. \ref{fig6}.

\textbf{In-kernel Client Behavior.} 
Upon successful establishment of an MPTCP connection, each client initializes and enters the \textit{Start} phase. During initialization, the client allocates metric buffers and spawns a dedicated kernel thread via workqueue to handle external communication. This design ensures that the sleepable operations remain in the process contexts instead of interrupt contexts of the kernel  \cite{bharadwaj2017linux}, preserving atomicity and non-preemption requirements for kernel-level operations. The Start phase employs a conservative slow-start mechanism similar to traditional algorithms to achieve initial rate ramping. It automatically terminates when the cwnd reaches a low threshold and sustains stable ACK reception, letting the agent assume primary control of decisions at the earliest viable stage.

Following the Start phase, the client transitions to the \textit{Train} phase and performs four functions: (i) continuous monitoring of per-ACK packet metrics such as delivered data and RTT; (ii) periodic queue draining operations to obtain propagation delay of the path; (iii) transmitting raw metrics to and receiving commands from external components through bi-directional IPC channels; (iv) enforcement of cwnd adjustments based on received directives.

To measure propagation delay, the client periodically enters a \textit{Probe} phase to capture uncontended RTT. It captures per-ACK data and RTT to assess current performance, using the maximum delivery rate over a recent window as a bandwidth estimate. Both idle delay and bandwidth estimates are managed with finite lifecycles, prompting periodic re-evaluation of path characteristics for long-term adaptation. External I/O is handled by the background thread that continuously transmits real-time observations from the main control loop and receives target cwnd directives. The main control loop incrementally adjusts cwnd per ACK, treating the target as an upper limit.

\textbf{User-space Proxy Behavior.}
The proxy acts as the intermediary, bridging the in-kernel clients with the external decision engine. It synchronizes the data from each subflow's client and aggregates these individual metrics into a time-aligned composite that represents a global snapshot of the entire MPTCP connection's health. To bridge the mismatched timescales between high-frequency ACK arrivals and the agent's decision-making interval, the proxy implements a short-term observation window. Within this window, it aggregates all incoming measurements to compute a smoothed representation of metrics (short-term average throughput and RTT). This mitigates the impact of transient network jitter to some extent, providing the agent with a more stable and reliable transmission representation. Specifically, the proxy also governs when to invoke the decision entity. While a single-step agent might require accumulating data across multiple windows to form a representative state, the sequence-aware model allows the proxy to trigger an inference after every single window, as the model itself leverages the historical sequence for robust decision-making. Conversely, when the agent issues a control decision, the proxy demultiplexes this command into specific cwnd directives for each subflow. They are then transmitted back to the respective kernel clients for enforcement.

\begin{figure}[t]
  \centering
  \includegraphics[width=0.85\columnwidth]{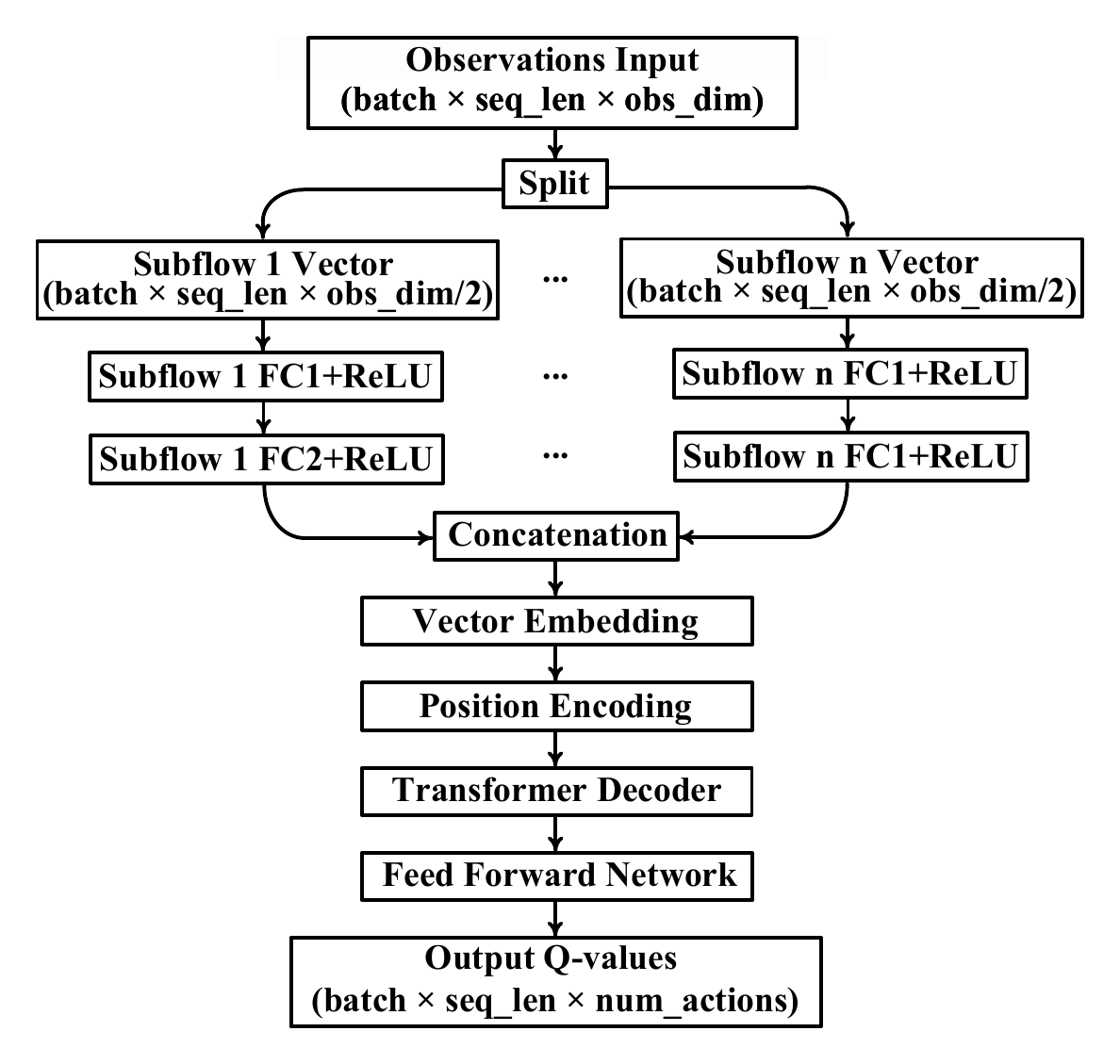}
  \caption
  {Network architecture of the Transformer-based Q-function in TCCO}
  \label{network structure}
  \end{figure}
  
\subsubsection{Transformer-based DRL Agent in External Decision Engine}
We propose a Transformer-based DRL agent to learn the non-linear interdependencies among subflow control actions and adapt to time-varying channel conditions without reliance on pre-defined models. Unlike single-step RL approaches that observe only the current state, our agent processes a sequence of historical observations, thereby mitigating partial observability. This sequential context enables the agent to learn a holistic control policy from experiential data, facilitating fine-grained, context-aware policy adjustments.

\textbf{Observation and Action Space.}
The agent operates on a context of the $L$ most recent observations, denoted as $\boldsymbol{c}_{t} = (o_{t-L+1}, \ldots, o_{t})$. As illustrated in Fig.~\ref{fig5}, each observation $o_t$ comprises per-subflow observations from all $M$ subflows. For each subflow $i \in \mathcal{W}$, the observation vector $o^i_t$ captures both instantaneous performance metrics and underlying path characteristics: the smoothed throughput, RTT, and cwnd. The physical properties of path are determined by estimating the bottleneck bandwidth from peak delivery rates and proactively measuring the base propagation delay through periodic, controlled queue draining. Additionally, an exploration flag, denoted $\mathit{expflag}$, is also embedded in the observation to facilitate proactive bandwidth exploration. The action space $\mathcal{A}$ comprises joint actions for all subflows. For subflow $i$, the action component $a^i_t$ specifies a discrete cwnd adjustment: $\Delta \mathrm{cwnd}_i = k \cdot \delta$, where $\delta \in \{j \mid j \in \mathbb{Z}, -n \le j \le n\}$, and $k$ is a scaling factor.

\textbf{Reward Shaping.}
The reward signal guides the agent toward efficient CC, which fundamentally involves deciding when to increase or decrease the cwnd to maximize transmission rates while maintaining acceptable queuing delay. The reward function $R$ is designed to teach the agent this timing for each subflow, using latency as an indicator of queue depth and incorporating the $\mathit{expflag}$ to prevent overly conservative policies.

The reward signal guides the agent toward efficient CC, which fundamentally involves deciding when to increase or decrease the cwnd to maximize transmission rates while maintaining acceptable queuing delay. Our reward function $R$ is designed to teach the agent this timing for each subflow, using latency as an indicator of queue depth and incorporating the $\mathit{expflag}$ to prevent overly conservative policies. Specifically, we adopt a hierarchical reward structure. First, boundary conditions are handled by deterministic rewards: a penalty of $-1$ is applied when the cwnd reaches its operational bounds or when the action contradicts the $\mathit{expflag}$. Conversely, if the $\mathit{expflag}$ is active and the action aligns with its indication, a reward of $+1$ is given. The $\mathit{expflag}$ is activated when no cwnd increase has occurred for a predefined period (6 time steps in our implementation), signaling the need to probe for available bandwidth. When no boundary condition is triggered, the agent receives a nuanced reward balancing throughput and latency. For subflow $i$, the reward is a weighted sum:
\begin{equation}
\label{eq:reward}
R_i = \alpha \cdot P_{D} + (1 - \alpha) \cdot R_{\rho}
\end{equation}
where $P_{D}$ is the RTT-related penalty and $R_{\rho}$ is the throughput-related reward. These reward components are based on the average of historical RTT $\overline{D}$ and throughput $\overline{\rho}$, yielding stable representations. The weighting factor $\alpha$ dynamically balances these objectives based on a threshold function $T(D_{\min})$ that determines when delay increases become detrimental:
\begin{equation}
\label{eq:threshold}
T(D_{min}) = \frac{\beta \cdot (1 + g \cdot (D_{min} - D_{f}) / \sigma)}{D_{min}}
\end{equation}
where $\beta$ defines the baseline permissible RTT, analogous to an acceptable queue depth. It can be tuned for specific network conditions and latency tolerances. The growth factor $g$ governs sensitivity to latency changes, $D_f$ provides a stable RTT floor, and $\sigma$ sets the RTT resolution such that fluctuations within $\sigma$s are treated equivalently.

The weighting factor $\alpha$ is computed via a sigmoid function to ensure smooth transitions:
\begin{equation}
\label{eq:alpha}
\alpha = \frac{1}{1 + e^{-\kappa \cdot (\frac{\overline{D}}{D_{min}} - T(D_{min}))}}
\end{equation}
where $\kappa$ controls the steepness of the transition. The RTT penalty is proportional to the excess delay ratio:
\begin{equation}
\label{eq:prtt}
P_{D} = -w_{D} \left( \frac{\overline{D}}{D_{min}} - T(D_{min}) \right)
\end{equation}
and the throughput reward is proportional to the average throughput:
\begin{equation}
\label{eq:rtput}
R_{\rho} = w_{\rho} \cdot \overline{\rho}
\end{equation}
where $w_D$ and $w_\rho$ are normalization factors ensuring comparable scales. The total reward aggregates contributions from all subflows: $R = \sum_{i=1}^{M} R_i$.

\textbf{Model Update.}
The agent's policy is derived from a Q-function $Q(\boldsymbol{c}, a)$, which estimates the expected cumulative reward for taking action $a$ in context $\boldsymbol{c}$. As shwon in Fig. \ref{network structure}, we employ a Transformer-based Q-function, which enables parallel computation in entire sequences. The agent follows an $\epsilon$-greedy policy: with probability $1 - \epsilon$, the agent selects the action maximizing the Q-value, i.e., $\pi(a \mid \boldsymbol{c}) = \arg\max_{a} Q(\boldsymbol{c}, a)$, while with probability $\epsilon$, a random action is selected to facilitate exploration. The Q-function is updated by minimizing the mean squared error:
\begin{equation}
\mathcal{L}(\theta) = \frac{1}{B \cdot L} \sum_{i=1}^{B} \sum_{t=1}^{L} \left( Q_\theta(\boldsymbol{c}_{i,t}, a_{i,t}) - y_{i,t} \right)^2
\label{eq:loss}
\end{equation}
where $\theta$ denotes the parameters of the Q-function, $B$ is the batch size, $L$ denotes the input sequence length, $\boldsymbol{c}_{i,t}$ and $a_{i,t}$ denote the context vector and the corresponding action at timestep $t$ of the $i$-th sample, respectively. The target Q-value is calculated using the Bellman equation:
\begin{equation}
y_{i,t} = r_{i,t} + \gamma (1 - d_{i,t}) Q_{\theta'}(\boldsymbol{c}_{i,t+1}, \arg\max_{a'} Q_\theta(\boldsymbol{c}_{i,t+1}, a'))
\label{eq:target}
\end{equation}
where $r_{i,t}$ denotes the reward of the $i$-th context sample, $d_{i,t} \in \{0, 1\} $ is an episode termination indicator, and $\theta'$ denotes the parameters of a periodically updated target Q-function that provides stable bootstrap targets during training.

\begin{figure}[t]
  \centering
  \includegraphics[width=\columnwidth]{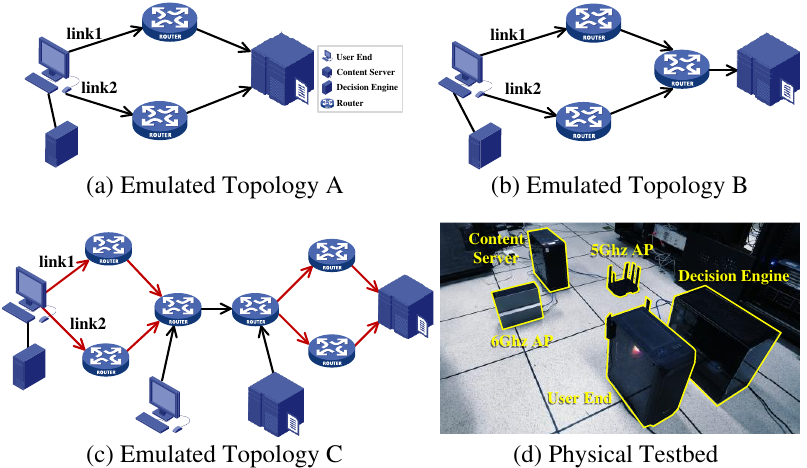}
  \caption
  {Topology of the experimental setup. The red line in (c) represents the link only traversed by multipath flows.}
  \label{fig7}
\end{figure}

\begin{table}[t]
\renewcommand{\arraystretch}{1.2}
\centering
\caption{Hyperparameters}
\begin{tabular}{ll}
\hline
\multicolumn{2}{c}{Network Hyperparameters} \\
\hline
Subflow $i$ FC1 \& FC2 dim & 64 \\
Embedding dim & 128 \\
Attention heads & 4 \\
Transformer layer & 1 \\
Positional Encoding & Learned \\
Gating & ResGate \\
Normalization & Post-LayerNorm \\
\hline
\multicolumn{2}{c}{Training Hyperparameters} \\
\hline
Gamma & 0.969 \\
Initial learning rate & $1 \times 10^{-4}$ \\
Minimum learning rate & $1 \times 10^{-6}$ \\
Decay rate & 0.96 (every 10 time steps) \\
Batch size & 32 \\
Replay buffer capacity & $1 \times 10^{6}$ \\
Target net update frequency & Every $1 \times 10^{4}$ time steps \\
Context length & 20 \\
\hline
\end{tabular}
\label{hyper}
\end{table}

\section{Evaluation and Validation}
\label{sec:validation}
We evaluated our method through extensive experiments conducted in both simulated and physical environments.  The simulation environment (Fig.~\ref{fig7}(a-c)) is implemented in Mininet with a modified Linux kernel version 5.4.243. For real-world validation, our physical testbed (Fig.~\ref{fig7}(d)) consists of a desktop equipped with Mediatek MT7922 and Intel AX210 NICs, establishing a dual-band (5GHz/6GHz) multipath WLAN connection to a storage server through TP-Link BE800 (6GHz) and ZTE AX5400Pro (5GHz) access points. Training and inference processes is accelerated using an NVIDIA RTX 4060 Ti GPU. The user device's kernel is upgraded to a modified Linux kernel version 6.8.0 to ensure driver compatibility with the NICs. We benchmarks TCCO against state-of-the-art MPTCP \cite{zhao2023multipath} and TCP \cite{vielhaus2025evaluating} implementations in both emulated and physical testbeds, while also assessing its performance across local and edge deployment scenarios. The hyperparameters is detailed in Table~\ref{hyper}.

\begin{figure}[t]
 \centering
 \includegraphics[width=\columnwidth]{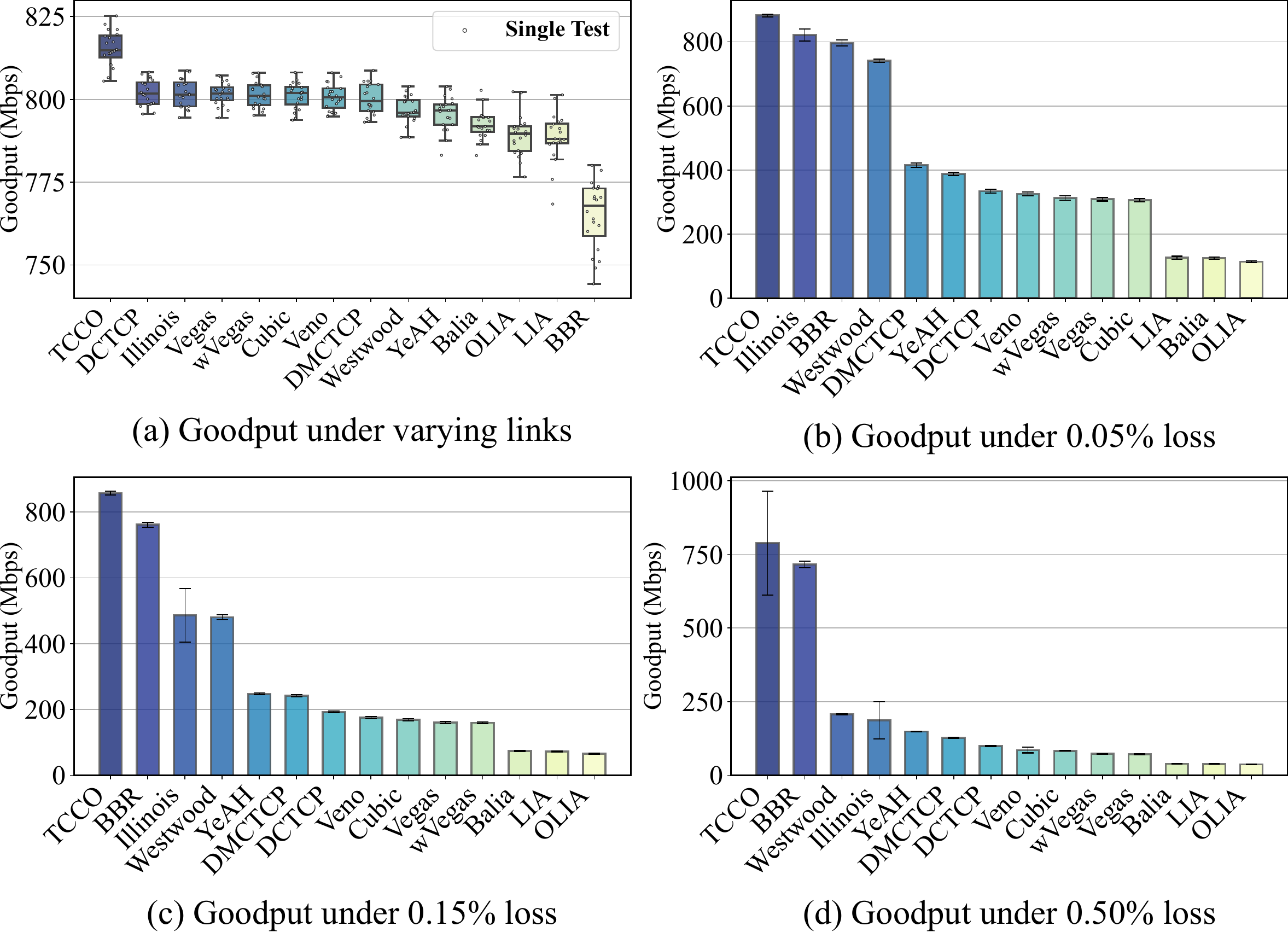}
 \caption
 {Average goodput comparison under varying links and different loss rates. The whiskers (a) and error bars (b-d) show the performance variation across multiple test runs.}
 \label{fig9}
 \end{figure}

\begin{figure*}[t]
\centering
\includegraphics[width=\textwidth]{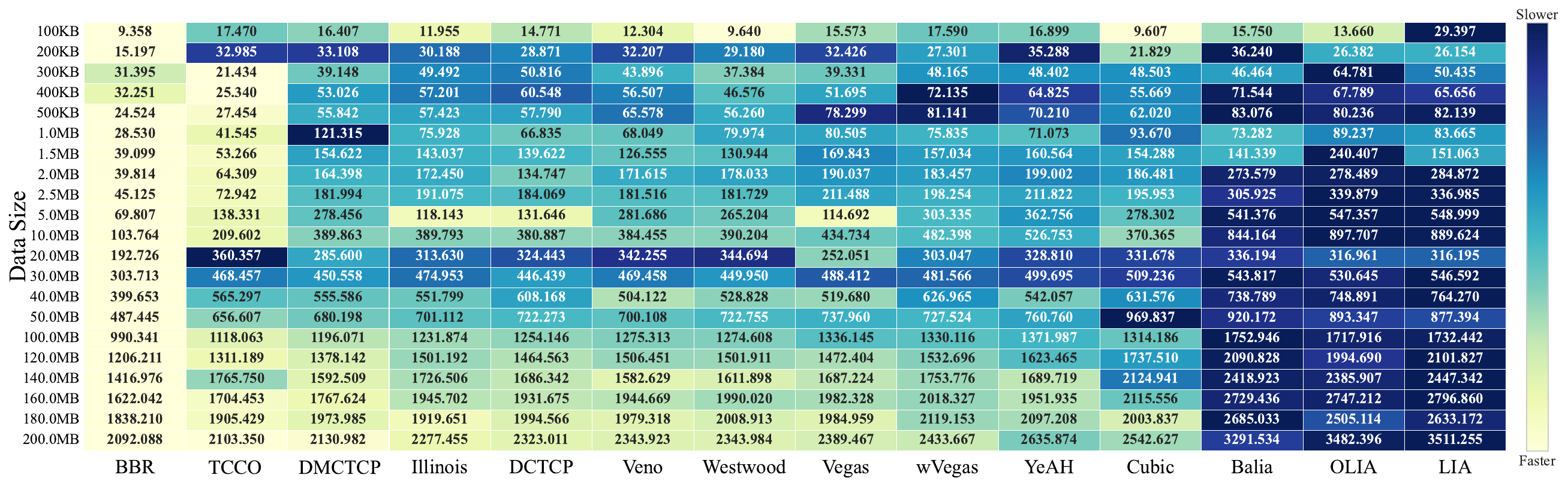}
\caption{Average FCT (ms) comparison under varying delay.}
\label{fig10}
\end{figure*}

\subsection{Evaluation in emulated environment}
We first evaluated bandwidth efficiency under fluctuating link capacity conditions, using the topology shown in Fig.~\ref{fig7}(a), which is identical to the setup described in Section~\ref{sec:motive}, Fig.~\ref{fig3}. In this configuration, each path's bandwidth oscillated between 400-500 Mbps with delay variations of 3-5 ms. Each algorithm underwent 20 tests to ensure statistical robustness against link variations. The results in Fig.~\ref{fig9}(a) shows that TCCO achieves the highest mean goodput at 815.47 Mbps. This represents a 1.75\% improvement over the next best MPTCP algorithm, wVegas (801.49 Mbps), and a 1.68\% advantage over the top-performing TCP algorithm, DCTCP (802.07 Mbps). Within the MPTCP algorithm group, TCCO surpassed Balia by 2.92\% and the TCP-friendliness-focused OLIA by 3.34\%. In the TCP group, DCTCP, Illinois, and Vegas formed the top tier. BBR's performance was the poorest, with its mean goodput of 765.01 Mbps being 6.2\% below its group average. This result should be interpreted with caution, as BBR's simulated performance diverged from real-world behavior.

Using the topology in Fig.~\ref{fig7}(a) where each path possesses 500 Mbps bandwidth, Fig.~\ref{fig9}(b-d) presents the performance under different packet loss rates, taking the mild loss scenario (0.05\% loss) as the baseline. TCCO showed robust resilience with only 2.8\% degradation at 0.15\% loss (882.40$\rightarrow$857.74 Mbps) and 10.7\% at 0.50\% loss (882.40$\rightarrow$787.96 Mbps). In contrast, loss-based algorithms suffered severe performance drops from the baseline: Cubic declined by 44.8\% (306.14$\rightarrow$169.00 Mbps) at 0.15\% loss and 72.8\% (306.14$\rightarrow$83.23 Mbps) at 0.50\% loss. Illinois showed similar vulnerability, dropping 40.9\% and 77.3\% respectively. BBR showed moderate resilience with 4.3\% and 10.1\% reductions.

To evaluate latency performance, we measure the Flow Completion Time (FCT) for short and medium-sized flows across various algorithms. The topology (Fig.~\ref{fig7}(a)) utilizes a 500 Mbps bandwidth per path with 1-3 ms fluctuating delay. Each algorithm undergoes 20 tests per flow size. The results in Fig.~\ref{fig10} reveal a nuanced performance landscape across different flow sizes. Overall, BBR consistently achieves the lowest FCT, and DTQN-based TCCO ranks second. For 100KB flows, TCCO's FCT is suboptimal, which can be attribute to the initial overhead of coordinating subflows that disproportionately impacts short transfers. For 1MB flows, as this overhead iss amortized, TCCO becomes competitive, outperforming most algorithms while trailing BBR. This trend continues for larger flows (100MB), where TCCO maintains its rank as the second-fastest method, surpassed only by BBR.

\begin{figure*}[t]
  \centering
  \subfloat[{\scriptsize Goodput under varied queue size}]{\includegraphics[width=0.248\textwidth]{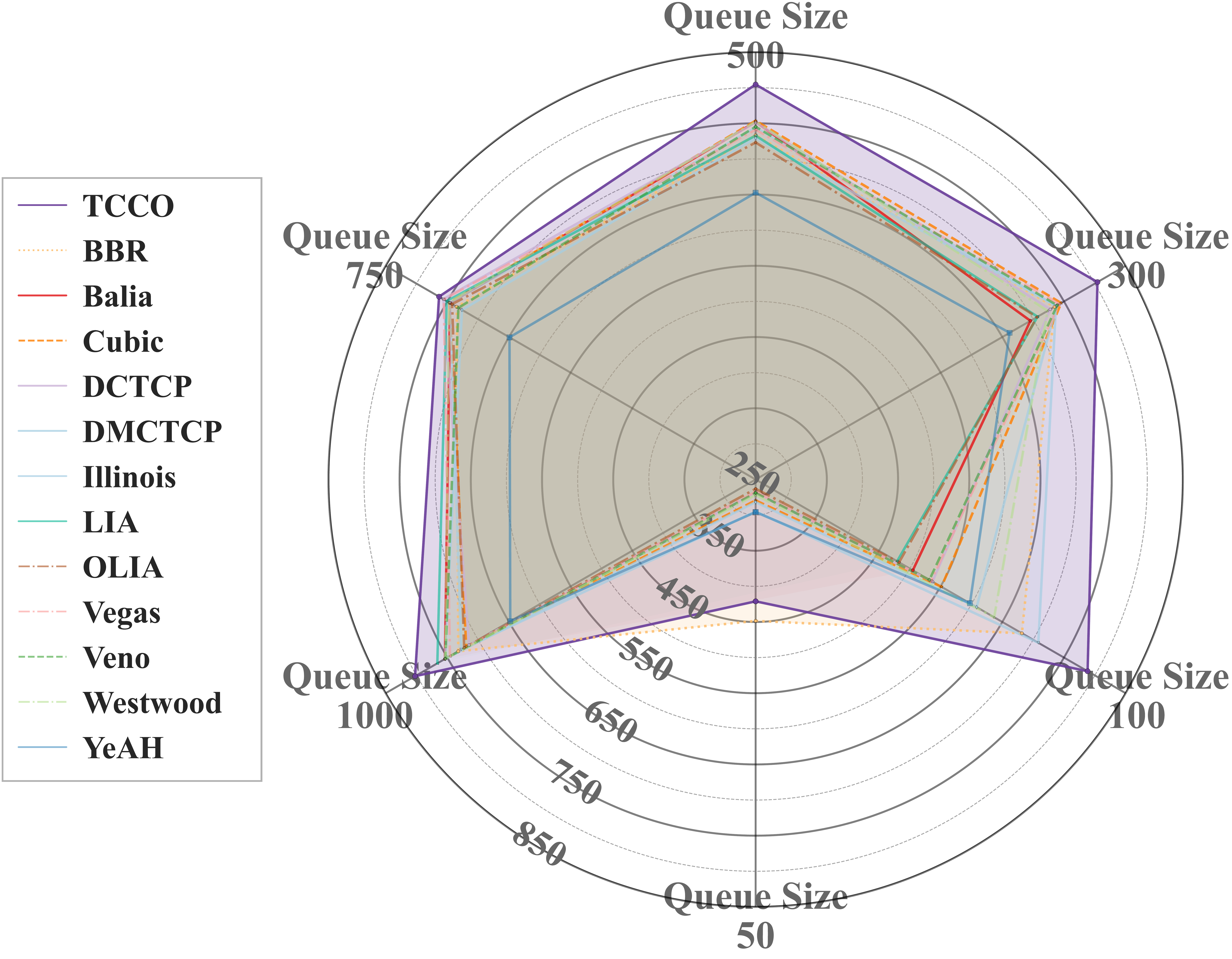}}\hfill
  \subfloat[\scriptsize FCT under 50 packets queue size]{\includegraphics[width=0.2505\textwidth]{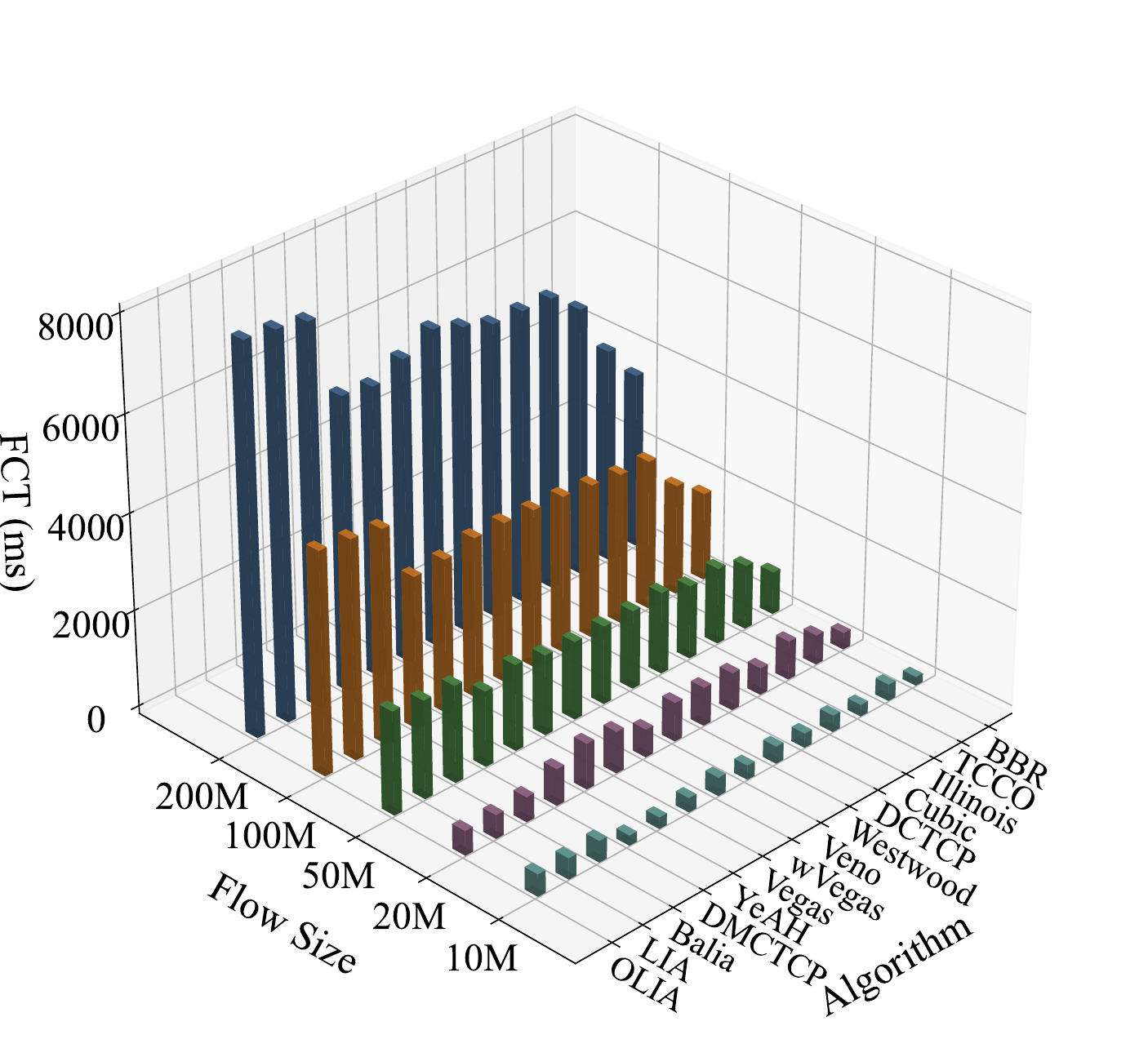}}\hfill
  \subfloat[\scriptsize FCT under 100 packets queue size]{\includegraphics[width=0.2505\textwidth]{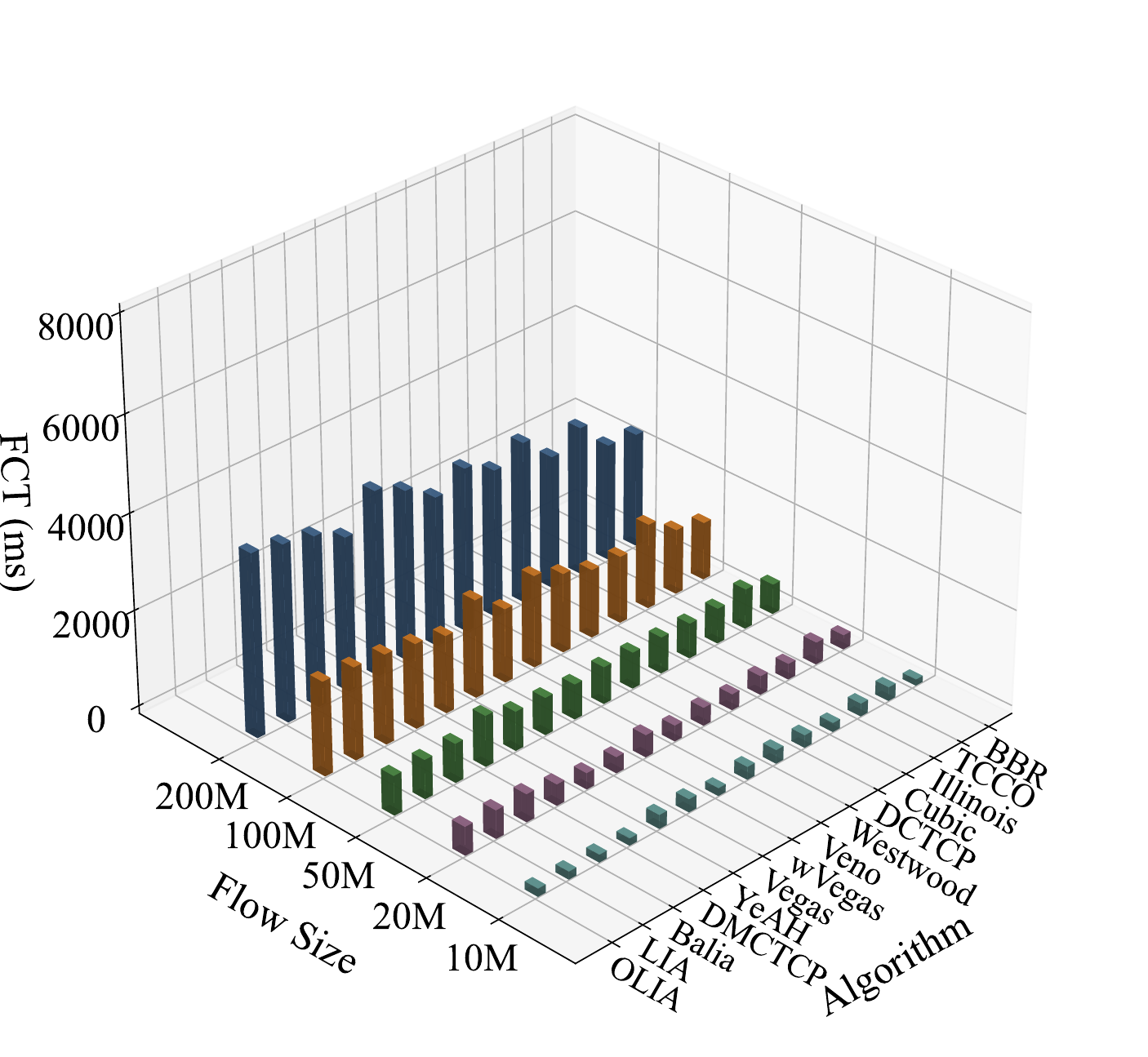}}\hfill
  \subfloat[\scriptsize FCT under 300 packets queue size]{\includegraphics[width=0.2505\textwidth]{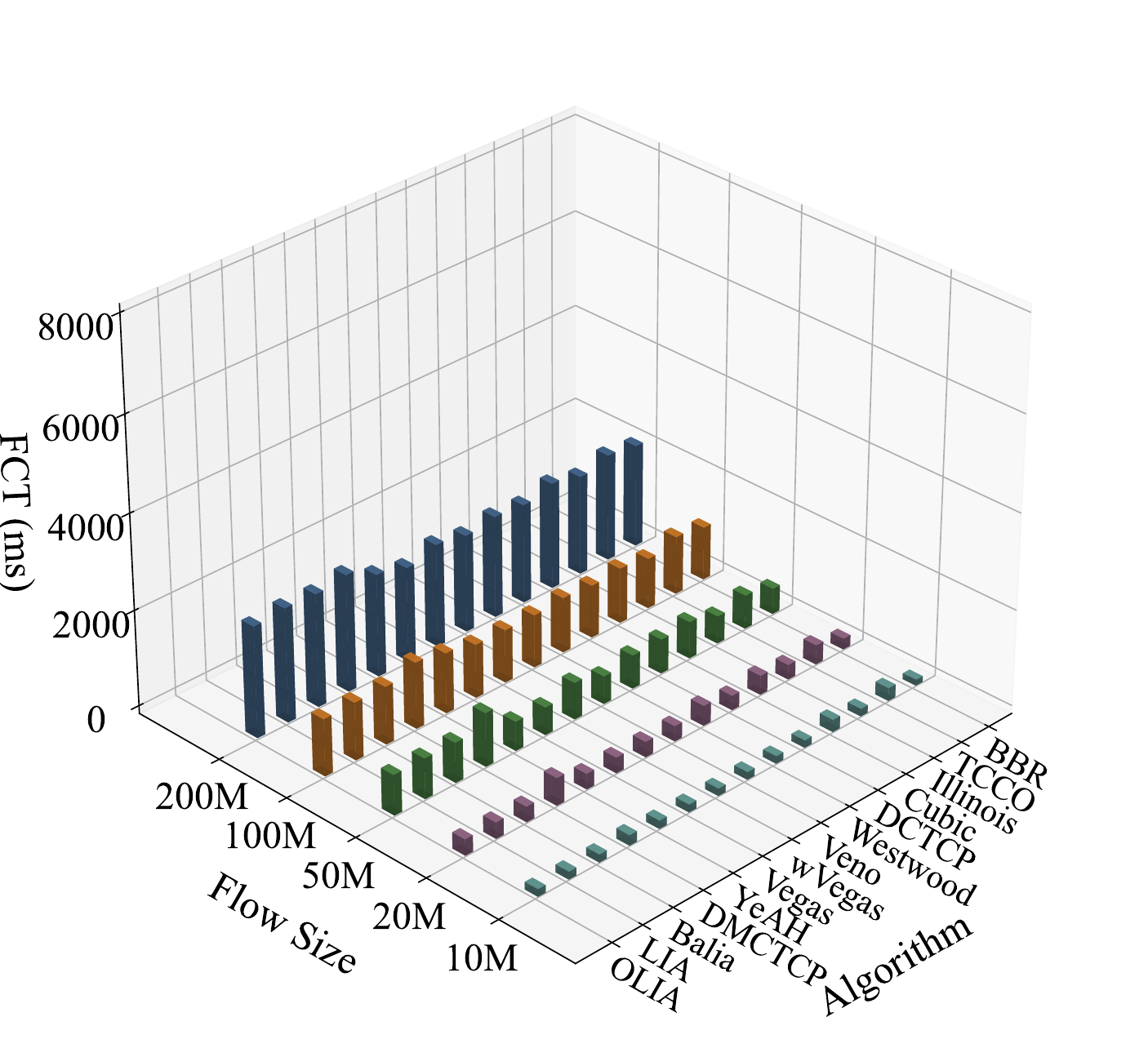}}\hfill
  \caption{Queue capacity impact on performance. The average goodput across different queue sizes is shown in (a), while (b-d) present FCT under queue sizes (we covered 10-200MB flows, given that queue constraints have little impact on shorter flows).}
  \label{fig11}
  \end{figure*}

% Overall, \textcolor{red}{TCCO showed competitive performance except for very short flows.}
We also investigate how each algorithm responds to varying buffer availability. Using the topology in Fig.~\ref{fig7}(b), we vary the bottleneck link's (1000 Mbps bandwidth) queue size to emulate buffer-to-BDP ratios ranging from 0.3 to 3, corresponding to buffer sizes of 50 to 500 packets (MTU of 1500 bytes). As shown in Fig.~\ref{fig11}(a), TCCO exhibits low sensitivity to buffer size, achieving 97\% of its peak throughput (804.52 Mbps) with only 100 packets of buffering. While BBR performs best under the smallest buffer scenario (50 packets), TCCO outperforms all algorithms at 100 packets and maintains this advantage as buffer size increases. In contrast, other algorithms such as Balia, LIA, and CUBIC show strong dependence on queue capacity, achieving significant throughput gains with larger buffers but suffering poor performance in shallow queues.

\begin{figure*}[t]
\centering
\includegraphics[width=2\columnwidth]{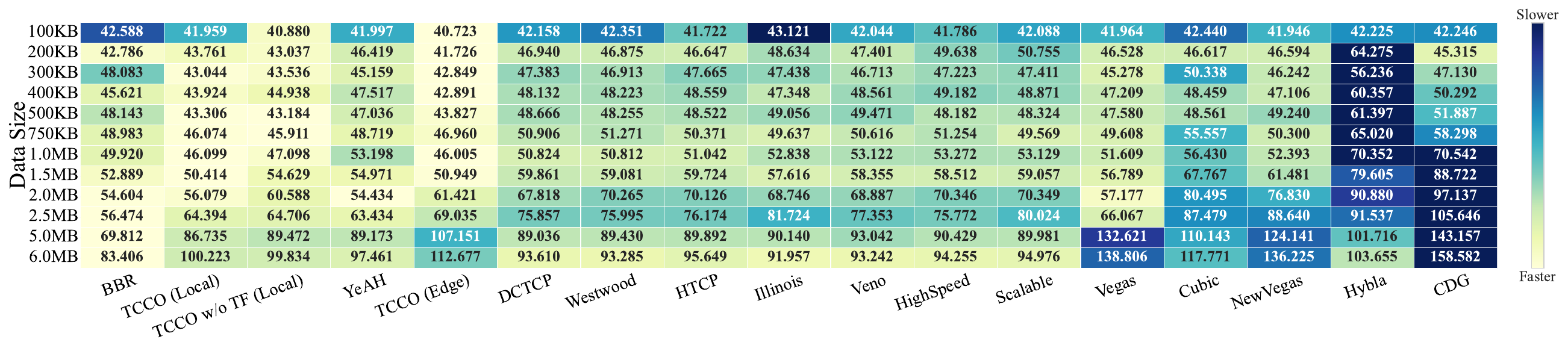}
\caption
{Average FCT (ms) comparison in physical testbed.}
\label{fig13}
\end{figure*}

\begin{table}[t]
\centering
\caption{Fairness of an MPTCP flow competing with an SPTCP flow, measured by Jain's Fairness Index (JFI).}
\label{tab:jfi}
\begin{tabular}{l|c|l|c|l|c}
\hline
\textbf{Algorithm} & \textbf{JFI} & \textbf{Algorithm} & \textbf{JFI} & \textbf{Algorithm} & \textbf{JFI} \\
\hline
YeAH & 0.9998 & Vegas & 0.9143 & BBR & 0.8408 \\
\hline
wVegaS & 0.9802 & BALIA & 0.9141 & TCCO & 0.8161 \\
\hline
DCTCP & 0.9732 & OLIA & 0.9136 & DMCTCP & 0.7802 \\
\hline
Westwood & 0.9554 & Illinois & 0.8945 & CUBIC & 0.7116 \\
\hline
Veno & 0.9373 & LIA & 0.8918 & & \\
\hline
\end{tabular}
\end{table}

Fig.~\ref{fig11}(b-d) illustrates FCT performance across varying queue capacities. Traditional algorithms like Balia, OLIA, and Westwood showed high buffer dependency, with FCT improvements of 61.9\%, 59.7\%, and 59.2\%, respectively, from 50 to 300 packet buffers. BBR and TCCO exhibited lower sensitivity, with improvements of 37.3\% and 39.6\%, indicating better adaptability to limited buffer environments. BBR excels with small buffers (50 packets) for large flows (200MB: 3649ms vs. TCCO's 4441ms) by avoiding buffer bloat. As buffer size increase to 300 packets, TCCO reduces the performance gap for large flows (2258ms vs. BBR's 2183ms, a 3.5\% difference) while maintaining strong performance across all flow sizes.

We evaluated TCP friendliness using the topology in Fig.~\ref{fig7}(c). As shown in Table~\ref{tab:jfi}, YeAH achieves the highest fairness (JFI: 0.9998). TCCO shows moderate fairness (JFI: 0.8161) as its design focuses on bandwidth efficiency optimization rather than TCP compatibility.

\subsection{Validation in Physical Testbed}

To validate our approach beyond simulations, real-world experiments are conducted using the dual-band Wi-Fi testbed shown in Fig.~\ref{fig7}(d). In addition to comparing TCCO against traditional TCP algorithms (other MPTCP congestion control algorithms are excluded from physical tests due to kernel/MPTCP version compatibility constraints), we specifically evaluate: (1) the performance difference between edge entity and local endpoint deployment, and (2) the comparison between sequence-based (TCCO) and single-step inference models (TCCO without the Transformer agent).

Fig. \ref{fig13} presents the average FCT comparison for short flows, where different algorithms show varying performance across flow sizes (each algorithm undergoes 20 tests per flow size). For the smallest flow (100KB), TCCO's performance is comparable to other algorithms (as for very short flows, cwnd adjustments are kernel-driven). For flows between 300KB and 1MB, locally deployed TCCO demonstrates superior latency performance, outperforming BBR by up to 10\% at 500KB (43.31ms vs 48.14ms). For larger flows (5-6MB), BBR proves effective, achieving FCTs approximately 20\% lower than TCCO. Edge-deployed TCCO shows moderate performance due to its millisecond-level control delay.

\begin{figure}[t]
  \centering
  \includegraphics[width=\columnwidth]{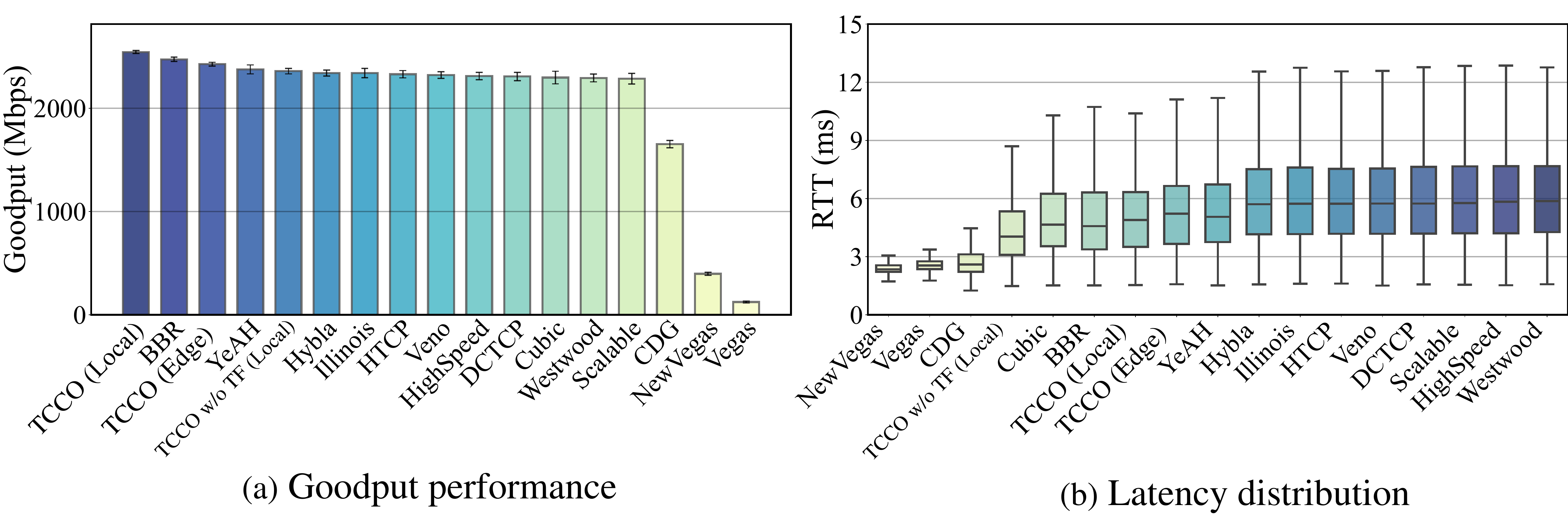}
  \caption
  {Goodput comparison in physical testbed. The error bars in (a) represent the standard deviation of the average goodput over multiple test runs, while the whiskers in (b) show the distribution of per-packet RTT.}
  \label{fig12}
\end{figure}

The goodput performance comparison in the physical testbed is presented in Fig.~\ref{fig12}(a). For each algorithm, at least 20 trials are conducted (each lasting 60 seconds), with half performed in a daytime office environment and the other half in an isolated server room at night (this setup applies to all subsequent experiments). TCCO demonstrates competitive performance under both deployment scenarios: local deployment achieves 2545.6 Mbps, while edge deployment achieves 2427.9 Mbps. The 4.8\% performance gap between local and edge deployment confirms that millisecond-level control latency introduced by remote decision-making does impact performance in high-speed multipath scenarios. However, this degradation is a reasonable trade-off given the gains in deployment flexibility and the ability to offload computations from resource-constrained edge devices. In comparison, TCCO without the Transformer agent only achieves 2360.4 Mbps.

Figure \ref{fig12}(a) presents the goodput comparison across different congestion control algorithms in our physical testbed. BBR achieves the highest throughput at 2474.9 Mbps among traditional algorithms. Hybrid algorithms such as YeAH (2377.3 Mbps) and Hybla (2342.0 Mbps) demonstrate decent performance by combining delay-based queue management with loss-based adaptation. Loss-based algorithms exhibit varied results: Illinois (2341.6 Mbps) performs well, while CUBIC (2298.4 Mbps) shows moderate effectiveness. Notably, the delay-gradient approach CDG (1652.7 Mbps) underperforms significantly in wireless multipath environments.

Fig.~\ref{fig12}(b) presents the per-packet RTT distribution. Delay-based algorithms exhibit the lowest RTT, with NewVegas (2.47ms mean, 0.48ms std) and Vegas (2.59ms mean, 0.39ms std) achieving minimal latency but at the cost of throughput. TCCO (deployed locally) maintains a balanced RTT profile (5.19ms mean, 2.23ms std) comparable to BBR (5.15ms mean, 2.40ms std), while delivering superior throughput. Edge-deployed TCCO shows a slight latency increase (5.45ms mean, 2.34ms std) due to control loop extension. The RTT stability, reflected in the standard deviation to mean ratio, is better in TCCO (0.43) than in BBR (0.47).

\begin{figure}[t]
\centering
\captionsetup[subfloat]{margin=0pt, captionskip=0.5pt}  
\subfloat[{Goodput Performance}]{\includegraphics[width=\columnwidth]{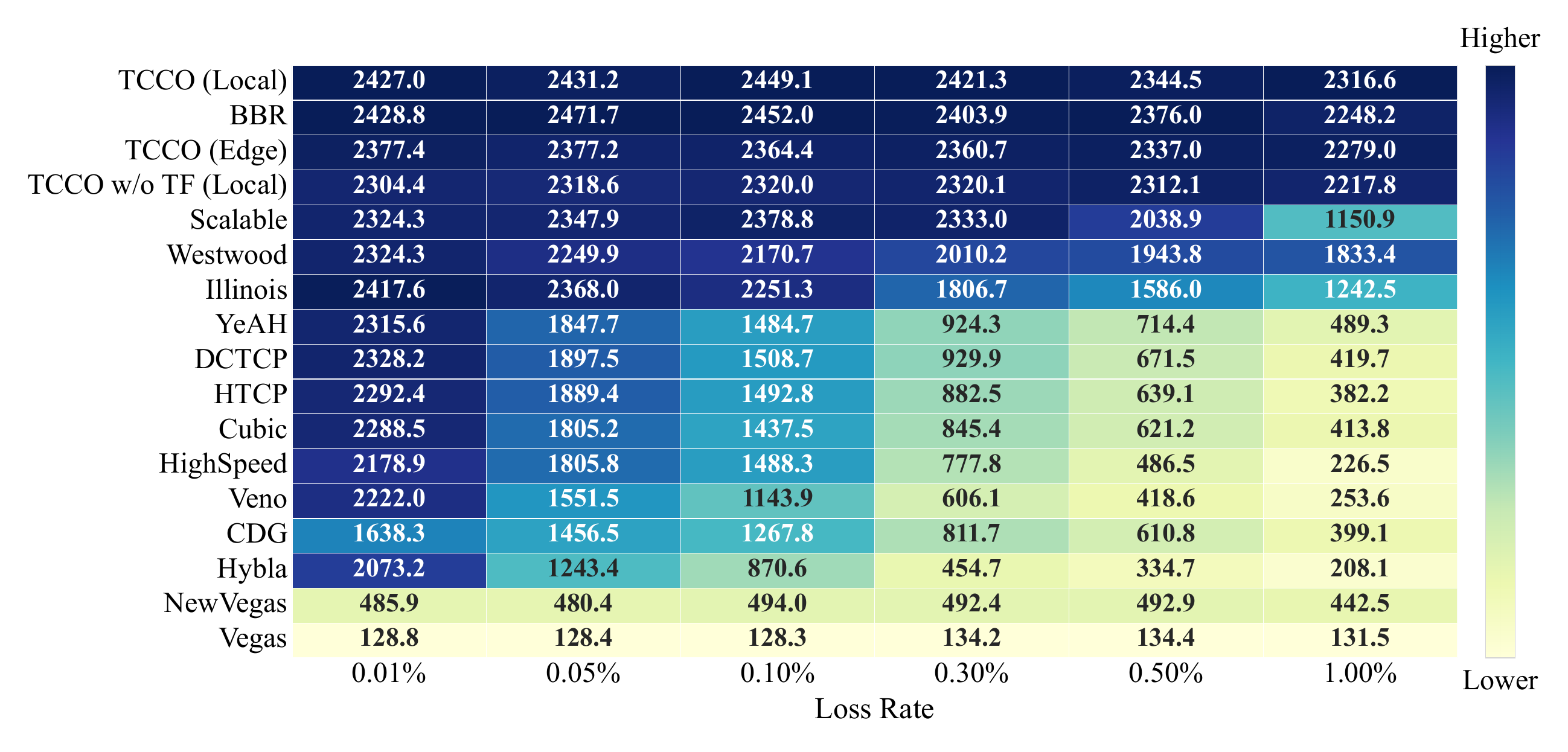}} \\
\subfloat[{Performance Degradation}]{\includegraphics[width=\columnwidth]{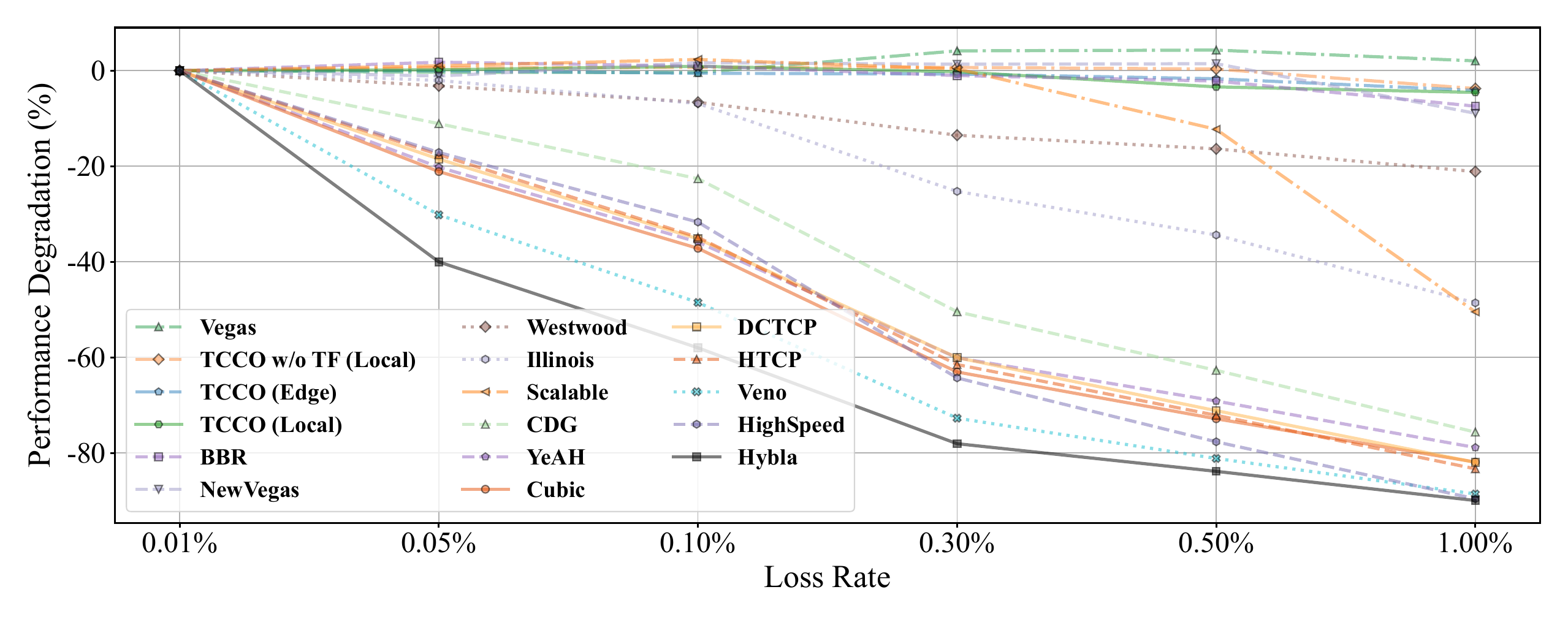}}
\caption{Goodput performance under varied stochastic loss. (a) shows the goodput (Mbps) of each algorithm at different loss rates. (b) shows the performance degradation (\%). Loss rates represent artificial loss introduced via Linux TC, excluding inherent link loss.}
\label{fig14}
\end{figure}

Fig. \ref{fig14} benchmarks the throughput robustness of various algorithms against stochastic packet loss. The TCCO family shows good robustness. At a 1\% loss rate, their performance degradation ranges from just 3.8\% to 4.5\%, while loss-based algorithms like CUBIC and HTCP experience throughput degradation of over 80\% under the same conditions. Locally deployed TCCO sustains an average throughput of 2316.56 Mbps in 1\% stochastic loss, outperforming BBR by 3.04\% and surpassing the average of loss-based algorithms (CUBIC, HTCP, and Illinois) by 283.53\%.

\section{Related Work}
\label{sec:related}
Non-learning-based CC algorithms use varied feedback and goals. Delay-based methods \cite{copa} proactively detect congestion through RTT increases, while loss-based algorithms \cite{ha2008cubic} \cite{illinois} react to packet drops as congestion indicators. ECN-based approaches \cite{dctcp, fmptcp} leverage explicit network feedback, primarily targeting data center environments with infrastructure support. Classic multipath CC algorithms aim to address fairness and efficiency. They extend single-path principles through coupled coordination \cite{lia, olia, balia}, linking subflow controls to balance TCP-friendliness, responsiveness, and achieve load balancing across heterogeneous paths \cite{wvegas}.

For better adaptability, many works have explored learning-based CC. Early attempts employ RL \cite{qtcp} and online learning \cite{pcc} to automatically adapt transmission strategies. DRL has further advanced CC design, with works \cite{drl3r, tcp-rl} showing improved performance in various environments. For multipath scenarios, RL is applied for latency-aware path management \cite{l2mptcp}. DRL has been extended to multi-agent frameworks for subflow coordination \cite{deepcc} and adapted for multipath optimization in specialized scenarios such as distributed edge learning \cite{fair_efficient_mptcp}. These learning-based approaches, while advancing the field, share a common limitation: their reliance on instantaneous state renders them unable to capture temporal patterns in network dynamics. A promising solution involves integrating sequence-aware architectures like RNNs \cite{drqn} and Transformers \cite{dtqn} \cite{dt} into DRL, which would enable agents to learn from historical data. Furthermore, the practical aspects of deploying deep-learning-based models to interact with the in-kernel datapath are often insufficiently discussed.

Emerging technologies enable datapath extensibility and flexibility. RDMA-based solutions \cite{redn} achieve complex network offloads through self-modifying chains on commodity NICs. DPDK \cite{dpdk_elettra} provides high-performance user-space packet processing with deterministic latency for real-time applications. The eBPF frameworks \cite{emptcp, frommgen2017programming} enable safe kernel extensibility through bytecode injection. QUIC's user-space implementation represents another departure from kernel-based datapath \cite{multiflow_quic}, though performance limitations over high-speed networks merit further investigation \cite{quic_performance}. 

% \subsection{From Heuristic Rules to Learned Policies}
% \subsection{The Challenge of State Representation in DRL}
% % \subsection{Learning to Control Network Congestion}
% \subsection{The Kernel Bottleneck for Network Intelligence}

\section{Conclusion}
\label{sec:conclusion}

In this paper, we propose TCCO, a Transformer-based Congestion Control Optimization framework designed to address the limitations of in-kernel multipath transport protocols in edge networks. By decoupling congestion control logic from the kernel datapath and offloading decision-making to an external external engine, TCCO enables flexible deployment of advanced, data-driven algorithms while maintaining compatibility with existing TCP/IP infrastructure. Leveraging a transformer-based DRL agent, TCCO effectively mitigates the impact of measurement noise, thereby enabling robust inference of network dynamics and coordinated control across multiple subflows. Extensive evaluation on both simulated and real-world testbeds demonstrates that TCCO achieves superior bandwidth efficiency and loss resilience, compared to state-of-the-art baselines. These results validate the practicality of TCCO for enabling a new class of rapidly deployable, edge-served network functions. As a potential future direction, we are looking forward to extending our method to improve the performance of various applications such as large language models~\cite{lin2023pushing,fang2026hfedmoe,zhang2025robust,lin2025hsplitlora,duan2025llm} and distributed learning system~\cite{zhan2025prism,hu2024accelerating,fang2026nsc,zhang2024satfed}.

\bibliographystyle{IEEEtran}
\bibliography{reference}

\end{document}